\documentclass[fleqn,usenatbib]{mnras}

\usepackage{amsmath}

\usepackage[T1]{fontenc}

\DeclareRobustCommand{\VAN}[3]{#2}
\let\VANthebibliography\thebibliography
\def\thebibliography{\DeclareRobustCommand{\VAN}[3]{##3}\VANthebibliography}

\usepackage{amssymb}	% Extra maths symbols
\usepackage{graphicx}	% Including figure files
\usepackage{txfonts}
\usepackage{titlesec}
\usepackage{multibib}
\usepackage[dvipsnames]{xcolor}
\usepackage{enumitem}
\usepackage{fancyhdr}
\usepackage[nointegrals]{wasysym} % astronomical symbols
\usepackage{subcaption}
\usepackage{orcidlink}

{% for the vertical padding

\title[Photoionisation of the WHIM by galaxy clusters]{Galaxy cluster photons alter the ionisation state of the nearby warm-hot intergalactic medium}

\author[L. {\v S}tofanov{\' a} et al.]{L{\' y}dia {\v S}tofanov{\' a},$^{1,2}$\thanks{E-mail: stofanova@strw.leidenuniv.nl}\orcidlink{0000-0003-0049-6205}
Aurora Simionescu,$^{2,1,3}$
Nastasha A. Wijers,$^{1,4}$
Joop Schaye,$^{1}$
and \newauthor Jelle S. Kaastra$^{2,1}$
\\
$^{1}$Leiden Observatory, Leiden University,
PO Box 9513, 2300 RA Leiden, The Netherlands\\
$^{2}$SRON Netherlands Institute for Space Research, Sorbonnelaan 2, 3584 CA Utrecht, The Netherlands\\
$^{3}$Kavli Institute for the Physics and Mathematics of the Universe (WPI), The University of Tokyo, Kashiwa, Chiba 277-8583, Japan\\
$^4$Center for Interdisciplinary Exploration and Research in Astrophysics (CIERA) and Department of Physics and Astronomy, Northwestern University,\\ ~ 1800 Sherman Avenue, Evanston, IL 60201, USA}

\date{Accepted 2022 June 29. Received 2022 June 29; in original form 2022 April 16}

\pubyear{2022}

\begin{document}
\label{firstpage}
\pagerange{\pageref{firstpage}--\pageref{lastpage}}
\maketitle

\begin{abstract}
The physical properties of the faint and extremely tenuous plasma in the far outskirts of galaxy clusters, the circumgalactic media of normal galaxies, and filaments of the cosmic web, remain one of the biggest unknowns in our story of large-scale structure evolution. Modelling the spectral features due to emission and absorption from this very diffuse plasma poses a challenge, as both collisional and photo-ionisation processes must be accounted for. 
In this paper, we study the ionisation by photons emitted by the intra-cluster medium in addition to the photo-ionisation by the cosmic UV/X-ray background on gas in the vicinity of galaxy clusters. For near massive clusters such as A2029, the ionisation parameter can no longer describe the ionisation balance uniquely. The ionisation fractions (in particular of \ion{C}{IV}, \ion{C}{V}, \ion{C}{VI}, \ion{N}{VII}, \ion{O}{VI}, \ion{O}{VII}, \ion{O}{VIII}, \ion{Ne}{VIII}, \ion{Ne}{IX}, and \ion{Fe}{XVII}) obtained by taking into account the photoionisation by the cosmic background are either an upper or lower limit to the ionisation fraction calculated as a function of distance from the emission from the cluster. Using a toy model of a cosmic web filament, we predict how the cluster illumination changes the column densities for two different orientations of the line of sight. For lines of sight passing close to the cluster outskirts, \ion{O}{VI} can be suppressed by a factor of up to $4.5$, \ion{O}{VII} by a factor of $2.2$, \ion{C}{V} by a factor of $3$, and \ion{Ne}{VIII} can be boosted by a factor of $2$, for low density gas.

\end{abstract}


%%%%%%%%%%%%%%%%%%%%%%%%%%%%%%%%%%%%%%%%%%%%%%%%%%

%%%%%%%%%%%%%%%%% BODY OF PAPER %%%%%%%%%%%%%%%%%%

\section{Introduction}
\label{sec:intro}

For almost three decades, scientists have been trying to find the `missing baryons' in the Universe. The `missing baryons' problem originates from the comparison of the amount of baryons detected in the high redshift Universe ($z>2$) with the amount detected in the local Universe from $z=0$ to $z=1-2$. At redshift $z>2$, these baryons reside in much cooler gas and can predominantly be detected in the so called Lyman $\alpha$ forest (e.g. \citealp{1971ApJ...164L..73L, 1980ApJS...42...41S, 1994ApJ...437L...9C, 1995ApJ...453L..57Z, 1997ApJ...490..564W, 2001ApJ...559..507S}). To predict where to find them in the local Universe, one needs to rely on cosmological hydrodynamical simulations (e.g. \citealp{1994MNRAS.267...13B, 1999ApJ...514....1C}), which show that baryons are heated to higher temperatures, mainly via shock-heating due to the gravitational collapse and the hierarchical growth of structures in the Universe. In addition, processes such as supernova feedback, active galactic nuclei feedback, radiative cooling, or photoionisation can heat up this gas (e.g. \citealp{2012MNRAS.425.1640T}). Most of the diffuse baryons in the present epoch have not yet been converted into stars and can be found in groups of galaxies -- intragroup medium (IGrM), in the haloes of galaxy clusters -- intracluster medium (ICM), or in the space between them --  intergalactic medium (IGM). A subset of the IGM is the warm-hot intergalactic medium (WHIM) permeating large-scale structure filaments, while another component of the IGM is comprised of gas found in the haloes of galaxies – the so-called circumgalactic medium (CGM).

In this work we focus on the WHIM, which at low redshift contains around $30$\% up to $60$\% of all baryons of the Universe (e.g. \citealp{1998ApJ...503..518F, 2001ApJ...552..473D, 2012MNRAS.425.1640T, 2012ApJ...759...23S, 2019MNRAS.486.3766M, 2021A&A...646A.156T}). These baryons reside predominantly in the filamentary structures of the cosmic web (e.g. \citealp{2019MNRAS.486.3766M, 2021A&A...646A.156T}) and can be detected mostly in the ultraviolet (UV, see e.g. \citealp{2003Natur.421..719N}) and X-ray wavebands/energies (e.g. \citealp{1998ApJ...509...56H, 2002ApJ...564..604F, 2019MNRAS.488.2947W, 2021arXiv210804847W}) because of its relatively high temperatures of $10^5 - 10^7$\,K. Due to its high temperatures, low electron densities ($10^{-6} - 10^{-4}$\,cm$^{-3}$), and high ionisation state, the observations of this very diffuse and tenuous gas are extremely challenging with currently available UV and X-ray missions. Since the emission decreases as density squared, and absorption decreases linearly with density, it is more feasible to detect the WHIM in absorption against very bright, point like sources (e.g. quasars). 
Since oxygen is the most abundant element after hydrogen and helium in Universe, it is common to look for the WHIM gas in \ion{O}{VII} and \ion{O}{VIII} absorbers, but also in \ion{Ne}{IX} or \ion{N}{VII} (e.g. \citealp{1998ApJ...503L.135P, 2003ApJ...586L..49F, 2003ASSL..281..109R, 2004PASJ...56L..29F, 2005ApJ...629..700N, 2007ApJ...655..831T, 2018Natur.558..406N, 2021arXiv210912146A}). Observations of the WHIM in emission, however, are possible, though mostly through stacking methods to obtain higher signal-to-noise to distinguish the WHIM detection from the background. Since we focus on absorption studies in this paper, we only list few of the publications regarding the emission studies (e.g. \citealp{1999A&A...341...23K, 2002A&A...394....7Z, 2003A&A...410..777F, 2003A&A...397..445K, 2008A&A...482L..29W}). The IGM can also be detected in the UV. Typical UV absorbers are coming particularly from warm CGM and can be observed in e.g. \ion{O}{VI}, \ion{C}{IV}, \ion{N}{V} or \ion{Ne}{VI} (see e.g. \citealp{2011Sci...334..948T, 2014ApJ...792....8W}). For more details we refer to the CGM review paper by \citet{2017ARA&A..55..389T}.

The WHIM gas is typically modelled with collisional ionisation equilibrium (CIE) models (assuming the gas has only one temperature) and with photoionisation models that take into account the photoionisaton by the UV and X-ray background (e.g. \citealp{2005Natur.433..495N}). The sources of this background are known to include star forming galaxies and quasars. The modelling of this background can be very complex, and has been described in many previous works, including e.g. \citet{2012ApJ...746..125H, 2020MNRAS.493.1614F}. 

The presence of the photoionising radiation causes a suppression of the cooling rates in comparison with the CIE case. In a CIE plasma, the lighter elements such as hydrogen, carbon, and helium are the dominant coolants for temperatures below $\sim 10^{-2}$\,keV (for a plasma with roughly proto-solar metallicities). In highly photoionised gas, these elements are significantly less efficient coolants. The relative contributions of different coolants to the total cooling rate, however, also depend strongly on the shape of the spectrum of the ionising source.  The suppression of cooling rates in photoionised plasma leads to longer cooling times and can affect star/galaxy formation over time. This has been already noted by e.g. \citet{1985ApJ...297....1S, 1987Natur.326..455D, 1991MNRAS.253P..31B}, which show how quasars can ionise the gas in their surroundings and inhibit the formation of galaxies in their neighbourhood. This can propagate through time and even affect the large-scale structure seen in the galaxy distribution. This was also shown in  \citet{1992MNRAS.256P..43E} for the gas of primordial composition (H and He plasma) where the paper explores how the presence of photoionisation caused by the extragalactic UV background can inhibit the formation of dwarf galaxies, most prominently in the gas temperature range $10^4$--$10^5$\,K (see also \citealt{1996MNRAS.278L..49Q, 1996ApJ...465..608T}).  \citet{2009MNRAS.393...99W} showed how UV/X-ray radiation from galaxies and quasars can significantly suppress the cooling rates for gas enriched with metals as well and how this affects the gas with temperatures even up to $10^7$\,K.

In this paper, we explore how an additional source of photoionisation, in this case caused by the photons originating in galaxy clusters in addition to the photoionisation by cosmic UV and X-ray background, can change the ionisation balance of the WHIM. In Section\,\ref{sec:methods} we model the spectral energy distribution of three different cool-core galaxy clusters together with the cosmic UV/X-ray background, which serves as the ionisation source to the photoionisation model. In Section\,\ref{sec:case_study_A2029}, we describe the main changes to the ionisation state of the WHIM by focusing on the most massive galaxy cluster in our sample. We make the comparison to the other two, less massive clusters in Section\,\ref{sec:comparison_clusters}. In Section\,\ref{sec:column_density_calc} we use a simplified model of a filament and predict column densities in two different orientations: perpendicular and parallel to the line of sight, and provide a comparison to the column densities calculated for the photoionisation by the cosmic UV/X-ray background only. In Section\,\ref{sec:cooling_rates} we show how the cooling rates of the WHIM can be affected by an additional source of photoionisation from the galaxy cluster. And finally, Section\,\ref{sec:conclusions}  summarizes our main conclusions. Throughout the paper, we assume a cosmology with total matter density $\Omega_m = 0.3$, dark energy density $\Omega_{\Lambda} = 0.7$, radiation density $\Omega_r = 0$, curvature $\Omega_k = 0$ and Hubble constant $H_0 = 70$\,km/s/Mpc.

%%%%%%%%%%%%%%%%%%%%%%%%%%%%%%%%%%%%%%%%%%%%%%%%%%%%%%%%%%%%%%%%%%%%%%%%%%%%%%%%%%%%%%%%%%%%%%%%%%%%
%%%%%%%%%%%%%%%%%%%%%%%%%%%%%%%%%%%%%%%%%%%%%%%%%%%%%%%%%%%%%%%%%%%%%%%%%%%%%%%%%%%%%%%%%%%%%%%%%%%%
\section{Methods}
\label{sec:methods}

\subsection{Galaxy cluster selection}
\label{sec:}

For the purpose of our study, we select three relaxed cool-core clusters with different masses and temperatures (A$262$, A$1795$ and A$2029$). We chose these clusters to estimate the effect of the cluster emission on its surrounding medium for a range of cluster parameters. We summarize the main properties of these clusters as reported by \cite{2006ApJ...640..691V} in Table \ref{tab:properties_Vikhlinin_clusters}: redshift $z$, radius $r_{500}$  \footnote{$r_{500}$ denotes the radius of a sphere within which the mean overdensity is $500$ times the critical density of the Universe.} and observational average temperature $T_{\rm spec}$, which is obtained from the single-temperature fit to the cluster spectrum (without the central $70$\,kpc region). 

\begin{table}
		\centering       
		\caption{Redshift $z$, $r_{500}$ and $T_{\rm spec}$ of the clusters used in our study taken from \citet{2006ApJ...640..691V}.}
		\begin{tabular}{|l|c|c|c|c|}
			\hline
			Cluster & z & $r_{500}$ [kpc] & $T_{\rm spec}$ [keV]  \\ \hline
			A262 & 0.0162 & 650 $\pm$ 21 & 2.08 $\pm$ 0.06  \\ \hline
			A1795 & 0.0622 & 1235 $\pm$ 36 & 6.12 $\pm$ 0.05 \\ \hline
			A2029 & 0.0779 & 1362 $\pm$ 43 & 8.47 $\pm$ 0.09  \\ \hline
		\end{tabular}
		\label{tab:properties_Vikhlinin_clusters}
\end{table}

\subsection{Density and temperature profiles}
Density and temperature profiles for the galaxy cluster sample used in our study are taken from \cite{2006ApJ...640..691V}. The emission measure profile $n_p n_e (r)$ follows
\begin{equation}
	n_p n_e{(r)} = n_0^2 ~ \dfrac{(r/r_c)^{- \alpha}}{\left( 1 + r^2/r_c^2 \right)^{3\beta - \alpha /2}} ~ \dfrac{1}{\left(1 + r^{\gamma}/r_s^{\gamma} \right)^{\epsilon/\gamma}} + \dfrac{n_{02}^2}{\left( 1 + r^2/r_{c2}^2 \right)^{3\beta_2} } \;,
	\label{eq:density_profile}
\end{equation}
where $n_p$ and $n_e$ are the proton and electron number density, respectively. Parameters $n_0, r_c, r_s, \alpha, \beta, \epsilon, n_{02}, r_{c2}$ and $\beta_2$ are taken from Table $2$ in \cite{2006ApJ...640..691V} and $\gamma = 3$. 

The temperature profile $T_{\rm 3D}{(r)}$ is expressed as a product of $t_{\rm cool}{(r)}$ and $t{(r)}$
\begin{equation}
	T_{\rm 3D}{(r)} = T_0 ~ \times ~ t_{\rm cool}{(r)} ~ \times ~ t{(r)} ;\,
		\label{eq:temperature_profile}
\end{equation}
where
\begin{equation}
	\begin{split}
		t_{\rm cool}{(r)} &= 	\dfrac{\left( \dfrac{r}{r_{\rm cool}} \right)^{a_{\rm cool}} + \dfrac{T_{\rm min}}{T_0}}{\left( \dfrac{r}{r_{\rm cool}} \right)^{a_{\rm cool}} + 1} \;,	\\
		t{(r)} &= \dfrac{\left(r/r_t \right)^{-a}}{\left[ 1 + \left( \dfrac{r}{r_t} \right)^b \right]^{c/b} } \;.
	\end{split}
\end{equation}
Parameters $T_0, r_t, a, b, c, T_{\rm min}, r_{\rm cool}$ and $a_{\rm cool}$ for individual clusters are taken from Table $3$ of \cite{2006ApJ...640..691V}. Both the emission measure profile and the temperature profile as defined by Eq.~(\ref{eq:density_profile}) and Eq.~(\ref{eq:temperature_profile}), respectively, are three-dimensional.

\subsection{SPEX }
\label{sec:spectrum}

In this study we use the SPEctral X-ray and EUV (SPEX) software package \citep{kaastra1996_spex, kaastra2018_spex, kaastra_j_s_2020_4384188} v. $3.06$\footnote{For the most recent version see \url{https://spex-xray.github.io/spex-help/changelog.html}} which is used for modelling and analysis of high-resolution X-ray spectra. With its own atomic database SPEXACT (The SPEX Atomic Code \& Tables) it includes around $4.2 \times 10^6$ lines from $30$ different chemical elements (H to Zn). More specifically, we use a model for collisional ionisation equilibrium \emph{cie} and the \emph{pion} model for photoionisation equilibrium (PIE) (described in \citealt{2016A&A...596A..65M}, for the most recent updates see \citealt{2021A&A...655A...2S}). Unless stated otherwise, we use the protosolar abundances by \citet{2009LanB...4B..712L} and assume that the intra-cluster medium (ICM) has solar metallicities\footnote{In practice, the contribution from line emission is subdominant to the bremsstrahlung continuum for the sources we study in this paper (hot and massive galaxy clusters, for which most of the emission comes from cluster cores).}.

\subsection{Total photon flux seen by an absorbing particle}
\label{sec:cluster_flux}

To calculate spectra of selected clusters, we treat them as extended sources. The clusters are divided into radial bins while setting the outermost radius to $r_{500}$ (we checked that the contribution to the cluster flux from $r_{500} < r < r_{200}$ is small in comparison with $r<r_{500}$). For the integration we use the \texttt{QUAD} function from the SciPy library \citep{2020SciPy-NMeth}, where the number of shells is set using an adaptive algorithm\footnote{This algorithm sets by default the upper bound of the number of subintervals in each integration step to $50$. The algorithm is described in detail at \url{https://docs.scipy.org/doc/scipy/reference/generated/scipy.integrate.quad.html}. We checked the integration for higher as well as lower number of subintervals and the change in comparison with the default number of $50$ subintervals is negligible (relative differences of order of $10^{-4}$ and lower for the comparison between $50$ and $10$ subintervals).}. For each shell at distance $r$ we calculate the density and temperature following Eq.~\ref{eq:density_profile} and Eq.~\ref{eq:temperature_profile}, respectively. Knowing the temperature of each shell we simulate the spectrum of the shell with the collisional ionisation equilibrium (CIE) model in SPEX. The normalization in SPEX is defined as the emission measure $\textnormal{EM} = n_e n_{\rm H}V$, where $V$ is the volume of the CIE source. Each CIE model is then renormalized by $n_p n_e$ following Eq.\,\eqref{eq:density_profile}.

\begin{figure}
	\centering
	\includegraphics[width=\columnwidth]{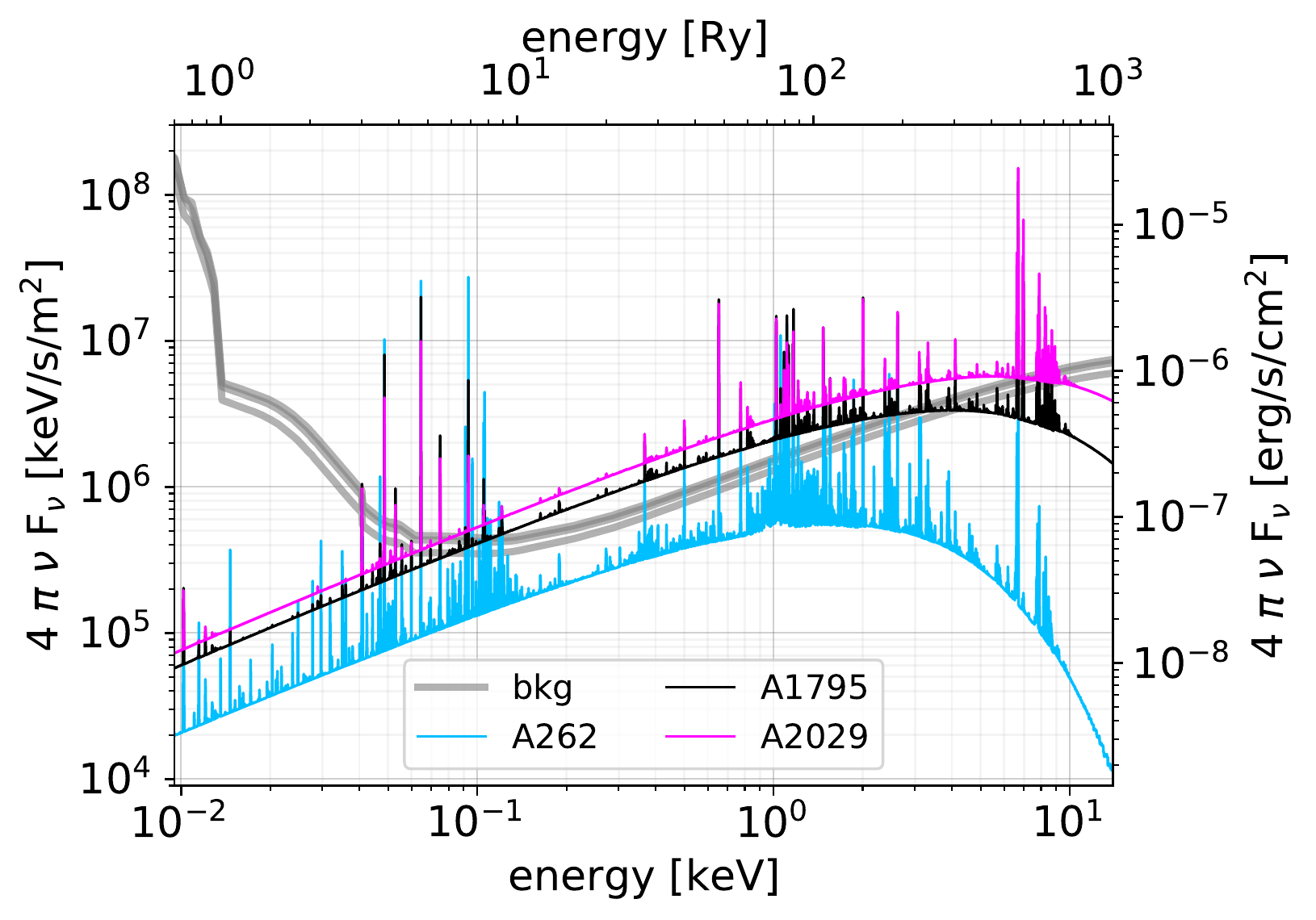}
	\caption{Spectra of the intra-cluster media of A$262$ (blue), A$1795$ (black) and A$2029$ (pink) and the corresponding cosmic UV/X-ray background at the cluster redshift (grey) as calculated by \citet{2020MNRAS.493.1614F}. We plot spectra for all clusters at a distance of $2 \times r_{500}$. The galaxy cluster spectra are treated as extended sources and their spectrum is a result of summation of CIE models of different temperatures (see the main text for more details).}
\label{Fig:SED_comparison}
\end{figure}

We estimate the photon flux seen by any absorbing particle for a set of distances $R_{\rm abs}$ between $r_{500}$ and $13$\,Mpc (for distances greater than $13$\,Mpc, cluster emission is negligible in comparison with the cosmic UV/X-ray background). The photon flux contribution of the infinitesimally small cluster shell (with a thickness $dr$) to the total photon flux can be written as the surface integral:
\begin{equation}
	\dfrac{dF_{\rm shell}^{\rm ph}}{dr} = \iint \dfrac{\varepsilon_r}{4 \pi\left( \vec{R}_{\rm abs} - \vec{r} \right)^2} ~ r^2 \sin{\theta} d\theta d\varphi  \;,
	\label{eq:F_surface_int}
\end{equation}
where $\vec{r}$\, is a vector from the cluster centre to the cluster shell and $\vec{R}_{\rm abs}$ is the vector from the centre of the cluster to the absorbing particle. $\mathit{\varepsilon}_r$ is the photon specific emissivity at given $r$. If we rewrite Eq.\eqref{eq:F_surface_int} by defining $L_{\rm shell} = \mathit{\varepsilon}_r \, 4 \pi r^2 dr $, the integral has an analytic solution given by
\begin{equation}
	F_{\rm shell}^{\rm ph} = \dfrac{1}{8 \pi r R_{\rm abs}	} L_{\rm shell}{(r)} \ln{ \left(  \dfrac{ R_{\rm abs} + r}{R_{\rm abs}-r} \right)  } \;,
	\label{eq:F_shell_analytic_solution}
\end{equation}
which holds for $R_{\rm abs} > r$. The total photon flux seen by an absorbing particle placed at distance $R_{\rm abs}$ from the galaxy cluster center is then the integral of Eq.\,\ref{eq:F_shell_analytic_solution} over radius $r$ from zero to $r_{500}$. 

The spectra of individual clusters are shown in Fig.~\ref{Fig:SED_comparison} for A262 (blue), A1795 (black) and A2029 (magenta) at a distance of $2\times r_{500}$. To account for the effect of photoionisation by unresolved background sources, we add the cosmic UV/X-ray background to the spectrum of the cluster. We use the model presented by \citet{2020MNRAS.493.1614F} (hereafter referred to as \emph{bkg}). For the redshifts of the cluster sample presented in this paper, the background does not differ much. Therefore, we plot all three profiles for the background in Fig.~\ref{Fig:SED_comparison} in grey. The final spectrum that we use as an ionising background in our calculations is then the sum of the cluster spectrum and the cosmic UV/X-ray background.

\subsection{Photoionisation model}
\label{sec:pion}
To model the effect of the galaxy cluster emission on the absorbing medium in its vicinity, we assume the medium is in ionisation equilibrium, accounting for collisional ionisation and photo-ionisation. We describe this plasma with the photoionisation model \emph{pion}. 

%We assume a spherical geometry in which a thin shell of the plasma surrounds the ionising source. 

In photoionised plasmas it is common to define the ionisation parameter $\xi$ \citep{1969ApJ...156..943T, 1981ApJ...249..422K} as
\begin{equation}
	\xi \equiv \dfrac{L_{\textnormal{1--1000 Ry}}}{n_{\rm H} R_{\rm abs}^2} \;,
	\label{eq:ionisation_par_xi}
\end{equation}
where the ionising source is described by the luminosity $L_{\textnormal{1--1000 Ry}}$ over the energy band $1$--$1000$ Rydbergs ($\approx$\,$1.36\times 10^{-2}$--$13.6$\,keV), $n_{\rm H}$ is the hydrogen number density of a photoionised plasma and $R_{\rm abs}$ is the distance of the photoionised plasma to the source of ionisation. 

It is common to tabulate ionisation fractions as a function of temperature and the ionisation parameter $\xi$. However, in the model that we are describing in this paper, this does not suffice. The reason is that the shape of the spectrum changes with the distance, because the relative contributions of different cluster shells and the relative contribution of the background all depend on the distance. This means that the ionisation balance can no longer be described solely as a function of $\xi$ and temperature, but it needs to be described as a function of $n_{\rm H}$, $R_{\rm abs}$ and temperature.

To account for this effect, we calculate $\xi$ based on a prescribed array of densities $n_{\rm H}$ and distances $R_{\rm abs}$ while taking into account the "correct" shape of the spectrum for each of these distances as described in Sec.\,\ref{sec:cluster_flux}. For that we select $30$ points for $n_{\rm H}$ and $30$ points for $R_{\rm abs}$ which are evenly distributed on a logarithmic scale. Densities $n_{\rm H}$ range from $10^{-6}$\,cm$^{-3}$ to $10^{-1}$\,cm$^{-3}$ and $R_{\rm abs}$ goes from $r_{500}$ to approximately $13$\,Mpc.  

In this paper, we use the \emph{pion} model in its temperature mode, which allows us to assume a range of temperatures for the photoionised plasma. This, however, means that instead of solving the ionisation balance and the energy balance equations simultaneously, we only solve  the ionisation balance equation. The consequence of this is that although we assume ionisation equilibrium, the plasma is allowed to be out of thermal equilibrium and the equilibrium temperature is only one of the temperatures in the range of the temperatures we used in our studies. This is reasonable for the gas we are probing, since this gas is shock-heated and tends to be out of thermal equilibrium because of its long cooling times. For our studies, we select $15$ different temperatures in the range $10^{-3}$--$1$\,keV ($\sim$ $10^4$--$10^7$\,K) evenly distributed on a logarithmic scale.

%%%%%%%%%%%%%%%%%%%%%%%%%%%%%%%%%%%%%%%%%%%%%%%%%%%%%%%%%%%%%%%%%%%%%%%%%%%%%%%%%%%%%%%%%%%%%%%%%%%%
%%%%%%%%%%%%%%%%%%%%%%%%%%%%%%%%%%%%%%%%%%%%%%%%%%%%%%%%%%%%%%%%%%%%%%%%%%%%%%%%%%%%%%%%%%%%%%%%%%%%
\section{Results}
\label{sec:results}

To demonstrate the effect of the additional source of photoionisation other than the cosmic UV and X-ray background, in subsection \ref{sec:case_study_A2029} we focus on the case of the cluster A2029. This cluster has the highest luminosity and alters the ionisation balance the most out of all selected clusters. In subsection \ref{sec:comparison_clusters} we show the comparison of A2029 to the less massive and less hot clusters A1795 and A262.

While we calculated the effect of the cluster emission for the entire grid of electron density, temperature, and $R_{\rm abs}$ as described in Section \ref{sec:pion}, for illustration purposes we present here our results for a limited, representative subset of these parameters.

\subsection{A2029}
\label{sec:case_study_A2029}

\subsubsection{The effect of the cluster emission on the total photoionisation and ionisation rates}
\label{sec:PIE_ion_rates}

\begin{figure}
	\centering
	\includegraphics[width=\columnwidth]{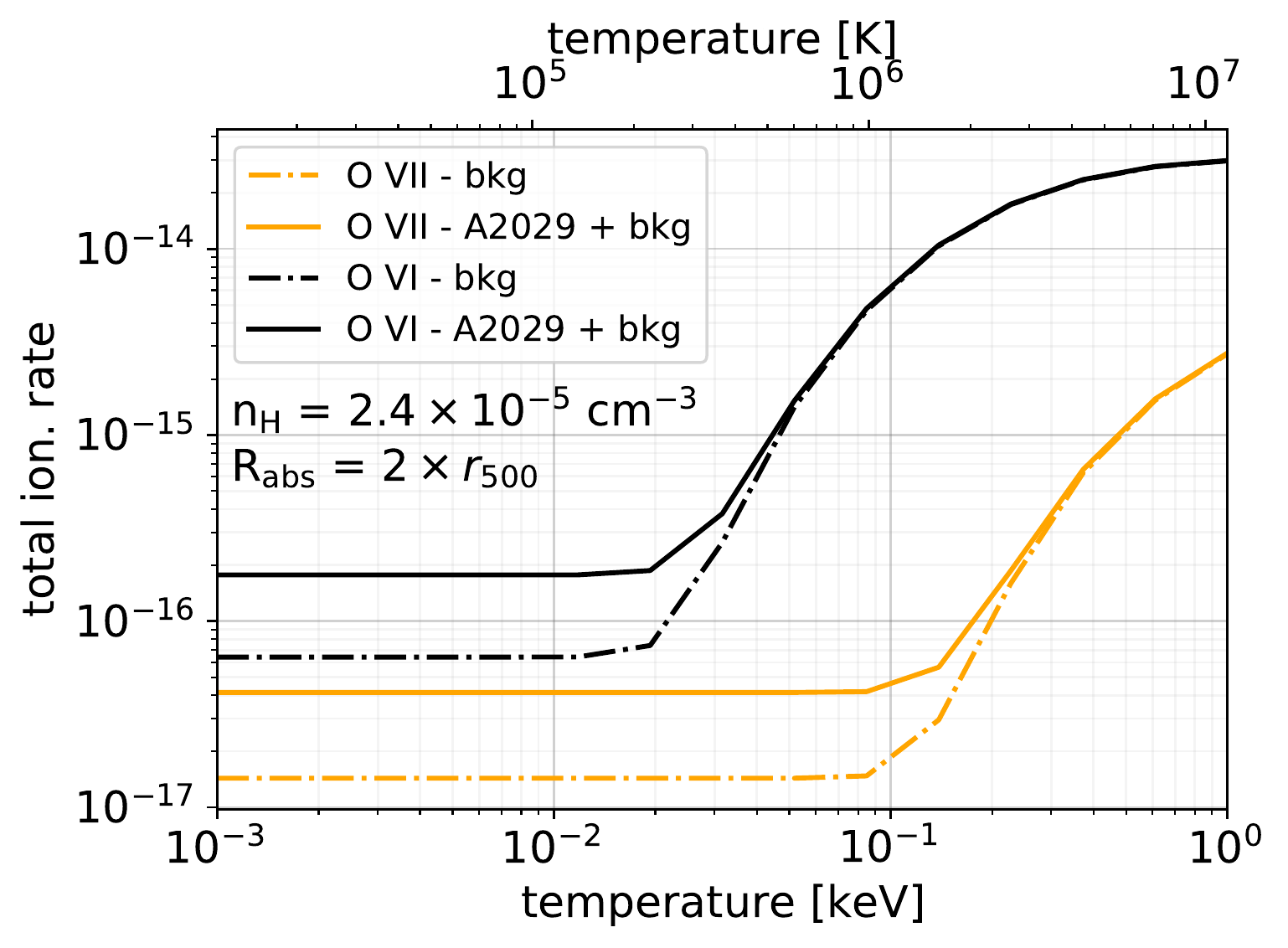} \\
	\includegraphics[width=\columnwidth]{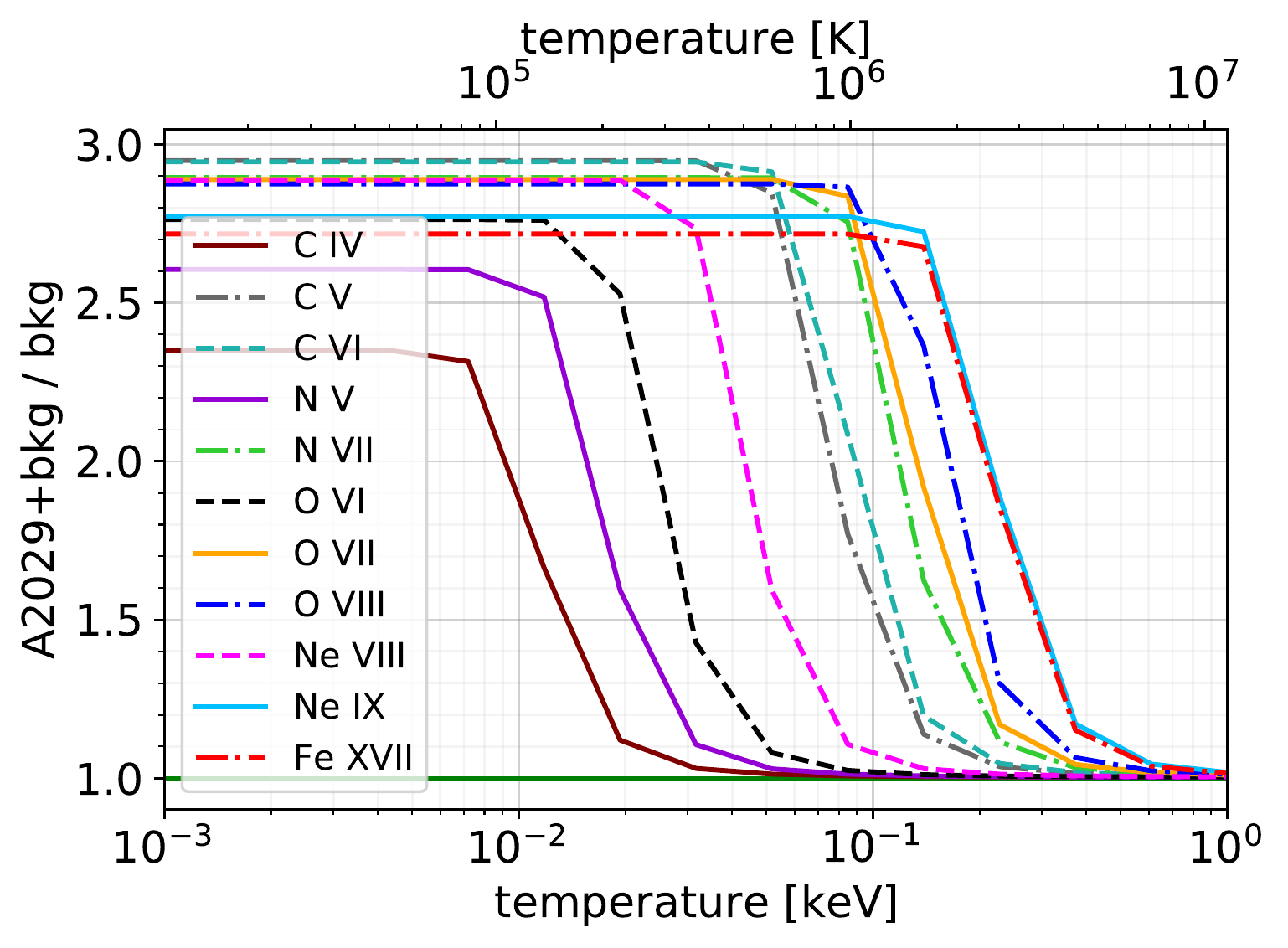}
	\caption{\textit{Top panel:} The total ionisation rate (number of  ionisations per second per ion) plotted as a function of photoionised plasma temperature for \ion{O}{VI} and \ion{O}{VII}. The dashed-dotted lines represent the total ionisation rate for the \citet{2020MNRAS.493.1614F} cosmic background at the redshift of cluster A$2029$ (z = 0.0779). The solid lines represent the total ionisation rate if the SED of A$2029$ plus the \citet{2020MNRAS.493.1614F} background is taken into account. The distance of the photoionised gas from the cluster center is $2\times r_{500}$ and the hydrogen number density is set to $2.4 \times 10^{-5}$\,cm$^{-3}$. \textit{Bottom panel:} ratio of the total ionisation rate for A2029+\emph{bkg} (solid lines in the top panel) and only the background \emph{bkg} (dash-dotted lines in the top panel) for a representative set of UV and X-ray ions. }
	\label{Fig:total_ion_rate_vs_kT}
\end{figure}

The top panel of Fig.~\ref{Fig:total_ion_rate_vs_kT} shows the total ionisation rate for the background spectral energy distribution (SED) in comparison to that for the A2029+\emph{bkg} SED as a function of temperature. The total ionisation rates are shown for the example of a hydrogen number density of $2.4 \times 10^{-5}$\,cm$^{-3}$ at a distance $2\times r_{500}$ of the photoionised gas from the galaxy cluster centre. The bottom panel of Fig.~\ref{Fig:total_ion_rate_vs_kT} shows the ratio of solid and dash-dotted lines in the top panel but for the example set of X-ray and UV ions: \ion{C}{IV}, \ion{C}{V}, \ion{C}{VI}, \ion{N}{V}, \ion{N}{VII}, \ion{O}{VI}, \ion{O}{VII}, \ion{O}{VIII}, \ion{Ne}{VIII}, \ion{Ne}{IX} and \ion{Fe}{XVII}. We note that the total ionisation and photoionisation rates are defined as number of ionisations/photoionisations per second per ion.

We can see from the plot that for low temperatures (below $ 20 $\,eV for \ion{O}{VI} and below $90$\,eV for \ion{O}{VII}) photoionisation is the dominant ionisation process. The addition of the cluster emission increases both the ionisation and the photoionisation rate. However, above a specific ion-dependent temperature, the plasma starts to be dominated by collisions with free electrons and the contribution of the cluster emission to the total ionisation rate decreases. At high temperatures, the plasma is in its CIE limit and the addition of the cluster emission does not have any effect on the total ionisation rate. The total photoionisation rate does not change as a function of $kT$, but it increases by a factor of few if A2029 is added to the background spectrum (e.g. by a factor of $2.8$ for \ion{O}{VI} and a factor of $2.9$ for \ion{O}{VII}). Since the WHIM gas may not always reach temperatures sufficiently high to be fully in CIE, taking into account the photoionisation from the galaxy cluster is important, mainly at lower gas temperatures, and should not be neglected.

To study the behaviour with the hydrogen number density, we compare plasmas close to the temperature of the CIE temperature peak ($kT \approx 0.2$\,keV) of \ion{O}{VII}, and approximately $10$ times lower than that ($kT \approx 0.02$\,keV). The total photoionisation rate does not change either with density, or with temperature (the photoionisation cross-section is a constant with temperature and density), it only changes with the distance to the cluster.

\begin{figure}
	\centering
	\includegraphics[width=\columnwidth]{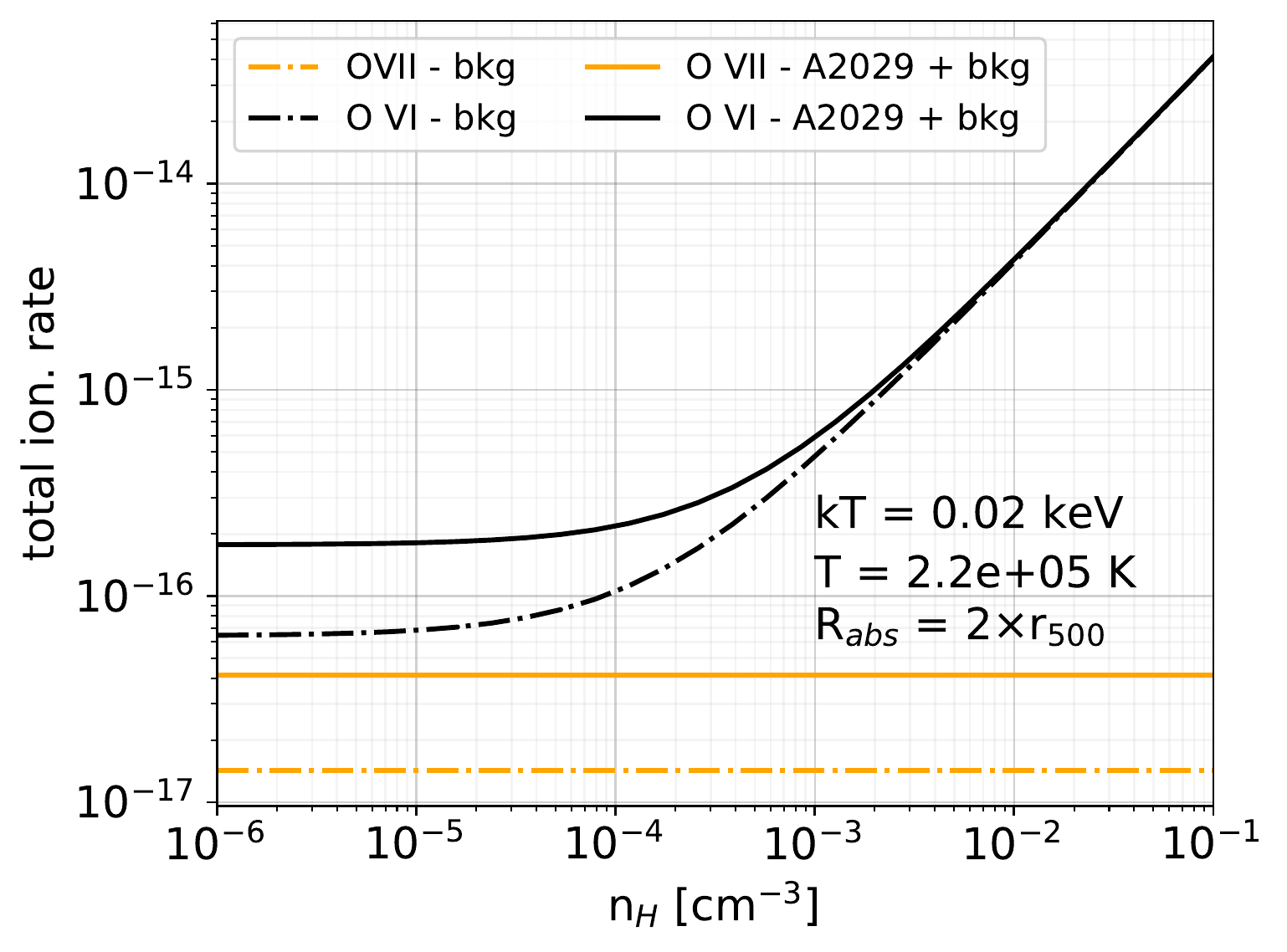} \\
	\includegraphics[width=\columnwidth]{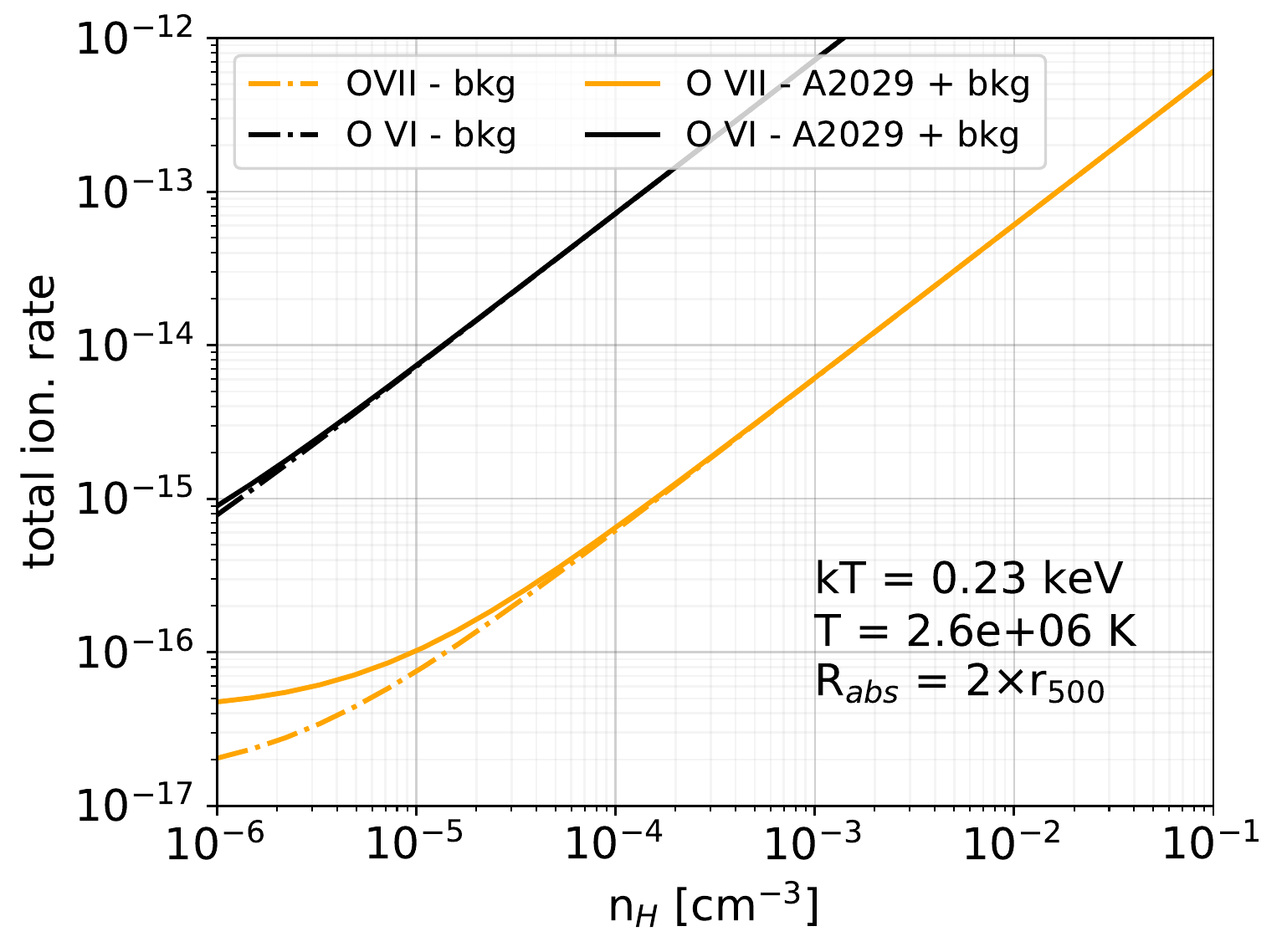} 
	\caption{The total ionisation rate (number of  ionisations per second per ion) plotted as a function of hydrogen number density for \ion{O}{VI} and \ion{O}{VII}. The dashed-dotted lines represent the total ionisation rate for the \citet{2020MNRAS.493.1614F} cosmic background at the redshift of cluster A$2029$ (z = 0.0779). The solid lines represent the total ionisation rate if the SED of A$2029$ plus the \citet{2020MNRAS.493.1614F} background is taken into account. The distance of the photoionised gas from the cluster center is $2\times r_{500}$. In the top panel we show plasma with temperature $0.02$\,keV (close to the CIE peak temperature of \ion{O}{VI}) and in the bottom panel we show plasma with temperature $0.23$\,keV (close to the CIE peak temperature of \ion{O}{VII}). }
	\label{Fig:total_ion_rate_vs_nH}
\end{figure}

The total ionization rates, plotted in Fig.\,\ref{Fig:total_ion_rate_vs_nH}, increase with density because of the increasing contribution from collisional ionisation, however, the details again depend on the temperature and the ion we study. As we see in the top panel of Fig.\,\ref{Fig:total_ion_rate_vs_nH}, at temperature $0.02$\,keV the total ionisation rate of \ion{O}{VI} is higher for A2029+\emph{bkg} in comparison with the background for densities $n_{\rm H} < 10^{-3}$\,cm$^{-3}$. As the density increases, the curves for A2029+\emph{bkg} and \emph{bkg}-only converge to the same value. However, for \ion{O}{VII}, the temperature is too low for collisional ionisation to contribute and all the ionisations come from photoionisation, which in this case is the same as when we described the behaviour of the photoionisation rate. For the higher temperature of $\approx 0.2$\,keV (bottom panel of Fig.\,\ref{Fig:total_ion_rate_vs_nH}), the addition of the cluster emission can generally be neglected.

\begin{figure}
	\centering
	\includegraphics[width=\columnwidth]{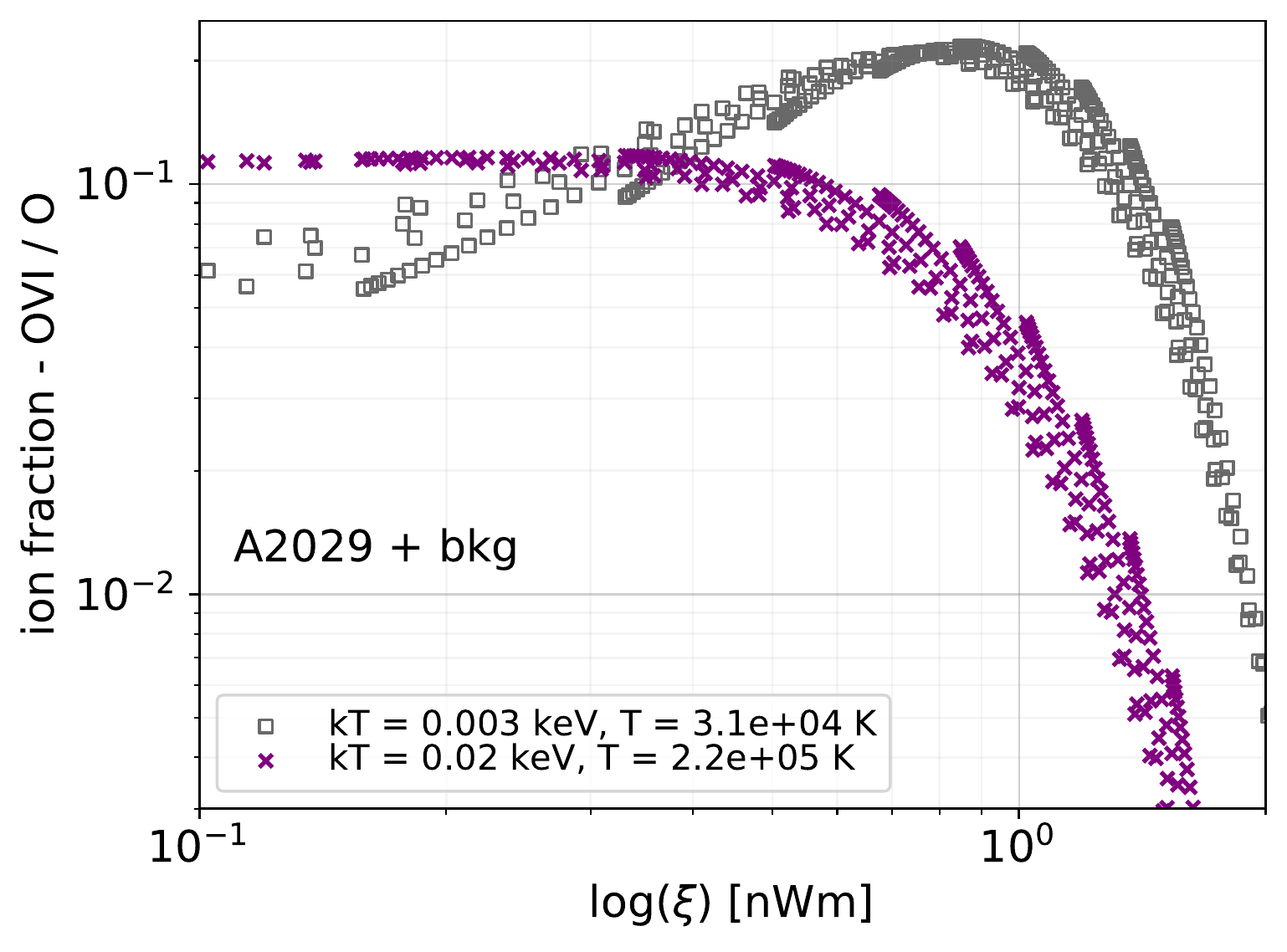}
	\caption{Ion fraction of \ion{O}{VI} as a function of the ionisation parameter $\xi$ for two chosen temperatures: $2.2\times 10^5$\,K (CIE peak temperature) and $3.1\times 10^4$\,K for a subset of distances $R_{\rm abs}$ and densities $n_{\rm H}$ selected in our studies. The figure illustrates how multiple ion fraction values are possible for a fixed value of $\xi$ as a consequence of including photoionisation by the SED of A2029+\emph{bkg}. }
	\label{Fig:ion_frac_vs_xi_A2029}
\end{figure}

In Eq.\eqref{eq:ionisation_par_xi} we defined the ionisation parameter $\xi$. As already mentioned in Sec.\,\ref{sec:pion}, it often suffices to parametrize the ionisation balance with this ionisation parameter and the temperature. However, in the model where we include the light from the cluster as well as the ionising background, the ionisation balance can no longer be parametrized solely with $\xi$ and temperature $T$, but needs to be expressed as a function of $(n_{\rm H}, R_{\rm abs}, T)$. We demonstrate this behaviour in Fig.~\ref{Fig:ion_frac_vs_xi_A2029}, where we plot ion fractions of \ion{O}{VI} as a function of the ionisation parameter $\xi$ for all $n_{\rm H}$ and $R_{\rm abs}$ values we used for the calculations and for two different temperatures. As Fig.~\ref{Fig:ion_frac_vs_xi_A2029} clearly shows, at a fixed value of $\xi$, many values of the ion fractions of \ion{O}{VI} are possible. The spread of possible values then depends on the density $n_{\rm H}$ of the ionising plasma and the distance $R_{\rm abs}$ from the ionisation source as well as its temperature $kT$. 

\subsubsection{The effect of the additional cluster emission on the ionisation balance}
\label{sec:ion_balance}
\begin{figure*}
	\centering
	\resizebox{\textwidth}{!}{
		\includegraphics{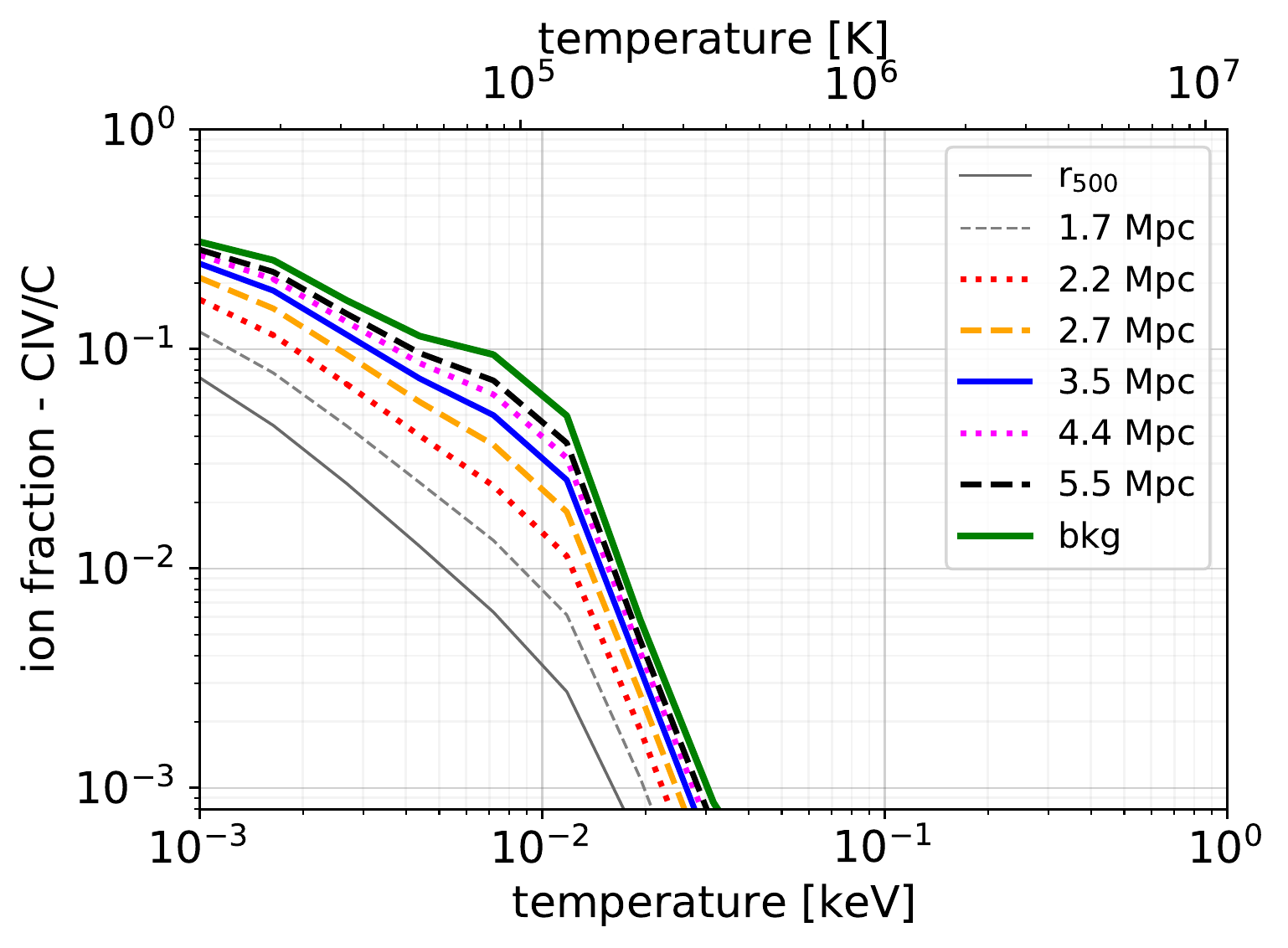}
		\includegraphics{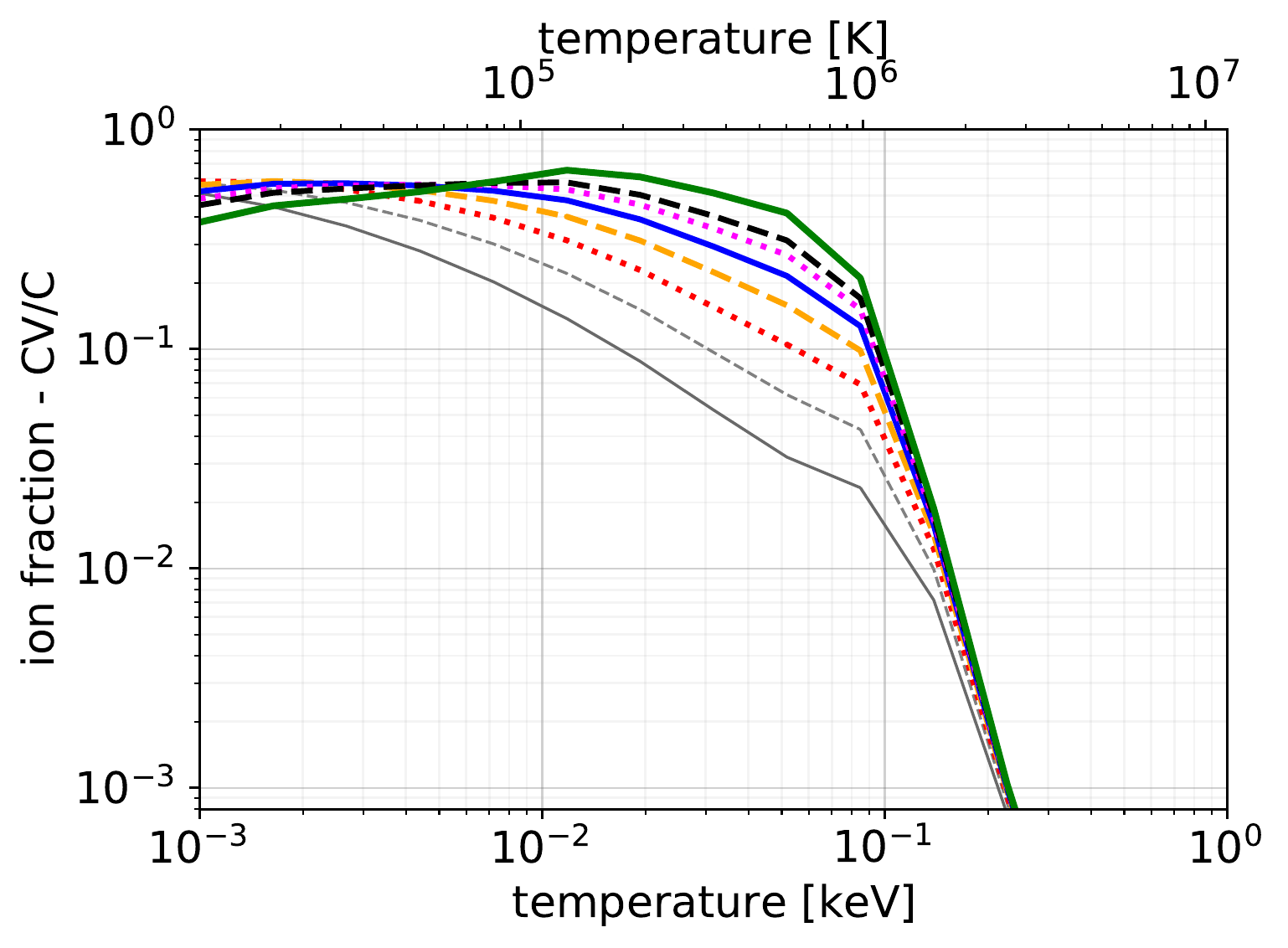} }
	\resizebox{\textwidth}{!}{
		\includegraphics{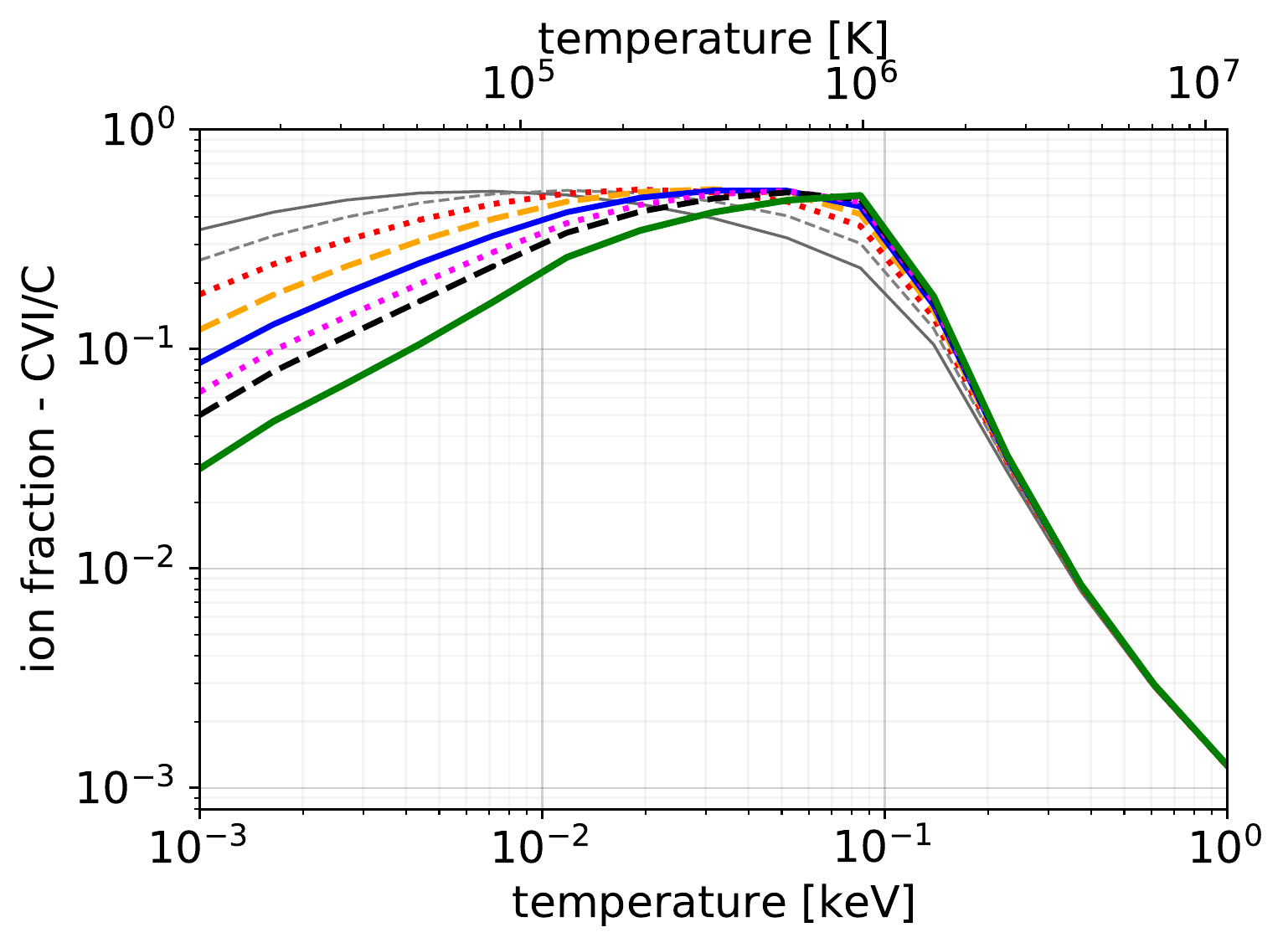}
		\includegraphics{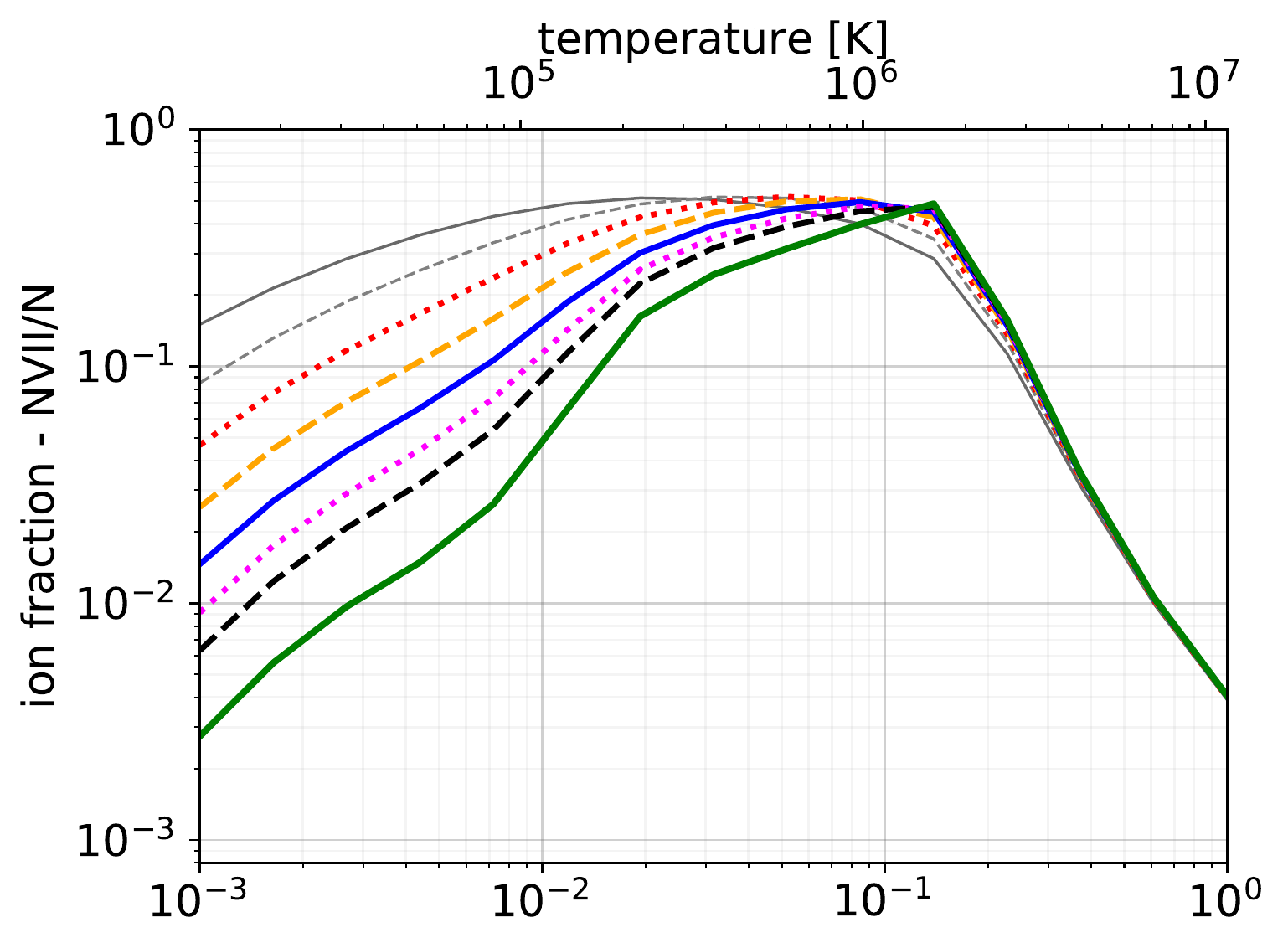} }
	\resizebox{\textwidth}{!}{
		\includegraphics{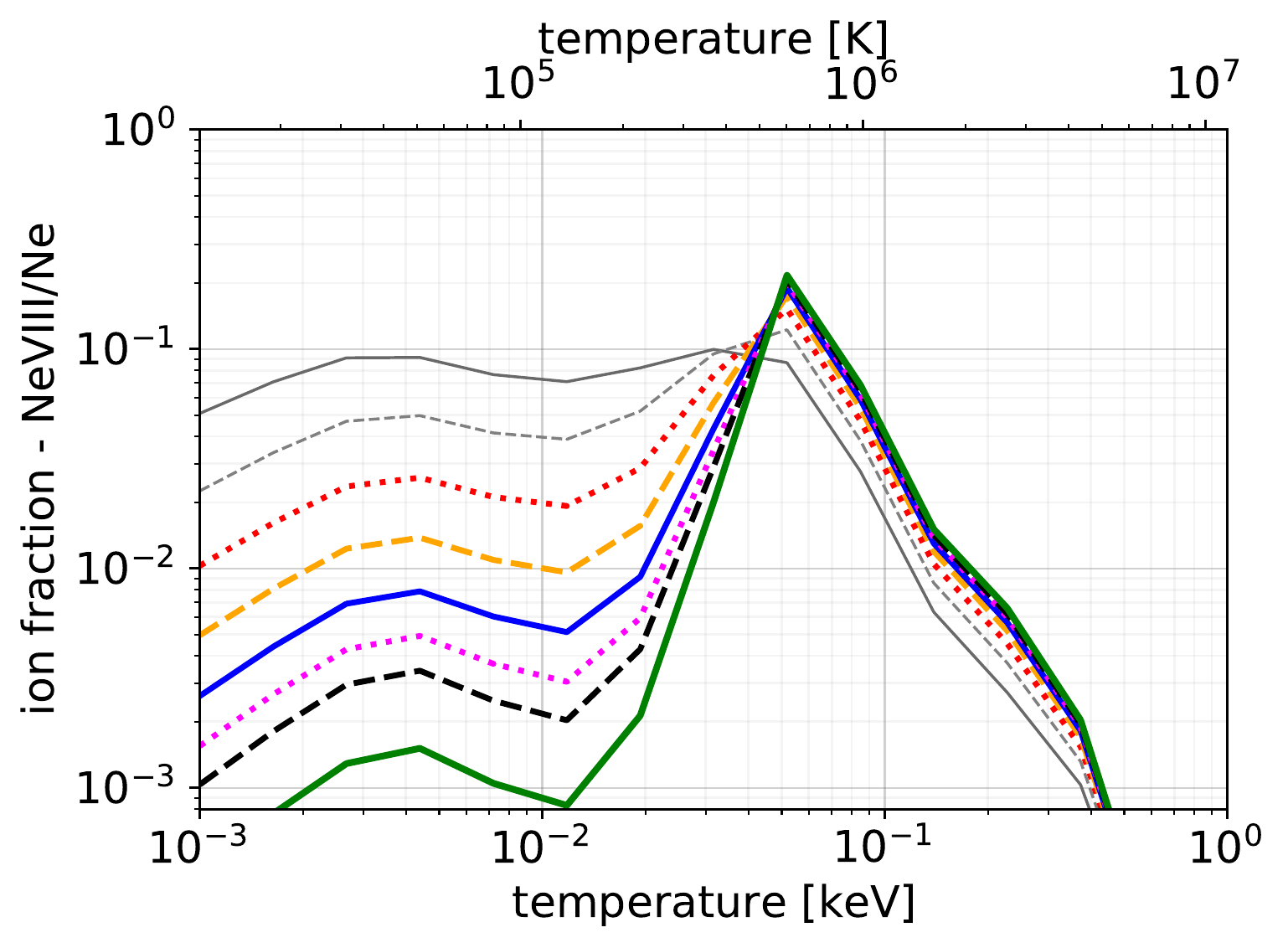}
		\includegraphics{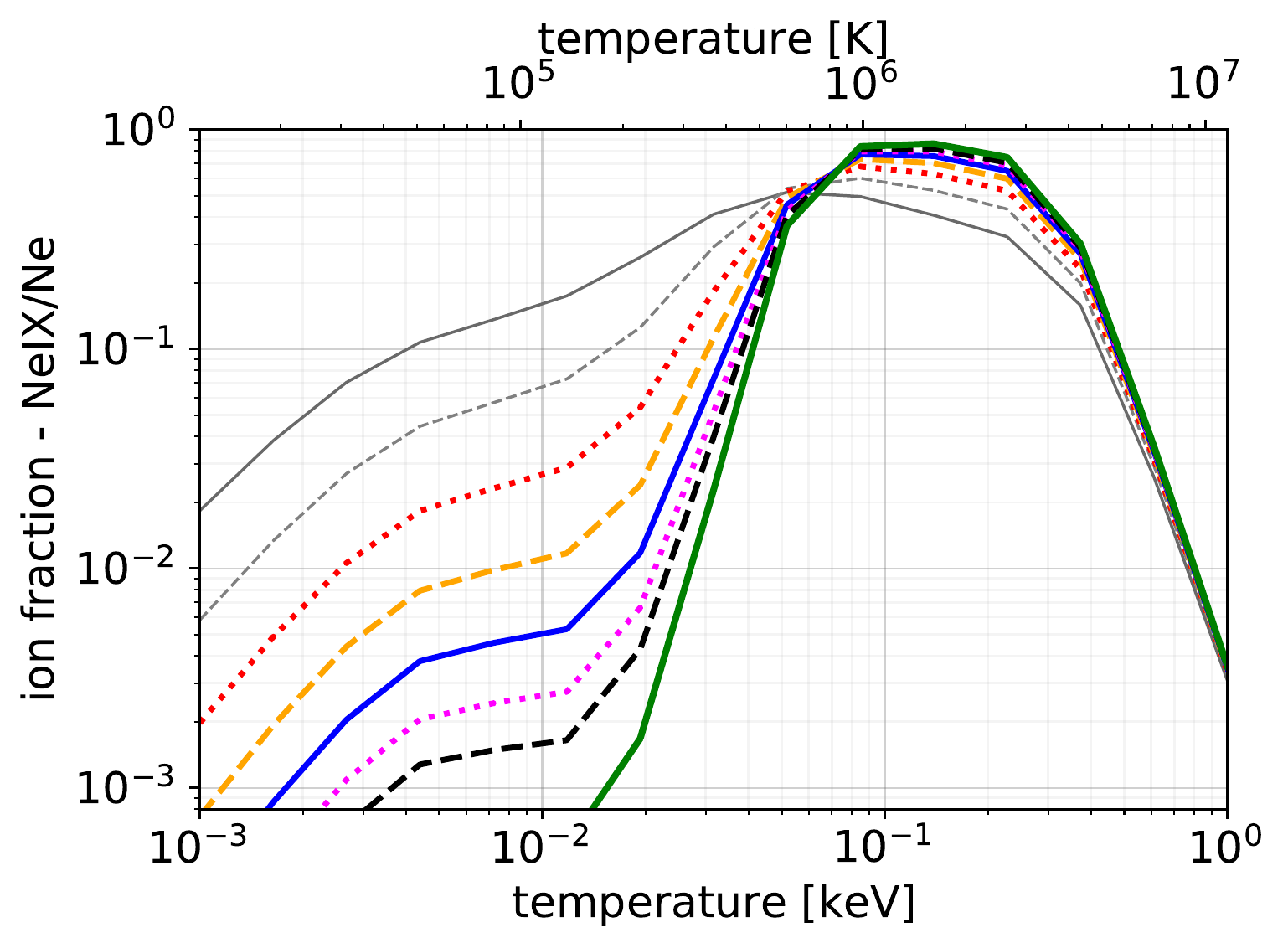} }
	\caption{Ion fractions of \ion{C}{IV}, \ion{C}{V}, \ion{C}{VI}, \ion{N}{VII}, \ion{Ne}{VIII}, \ion{Ne}{IX} as a function of plasma temperature for A2029+\emph{bkg} and the cosmic X-ray/UV background by  \citet{2020MNRAS.493.1614F}. Different colors represent ion fractions for different distances from the cluster center as indicated in the legend and the green solid line shows the ion fractions for only the background. The hydrogen number density is assumed to be $2.4 \times 10^{-5}$\,cm$^{-3}$. The ionisation parameters $U$ and $\xi$ which correspond to distances shown in this plot are listed in Table \ref{Table:U_xil_par}.}
	\label{Fig:ion_fractions_A2029_update}

\end{figure*}

\begin{figure*}
	\centering
	\resizebox{\textwidth}{!}{
		\includegraphics{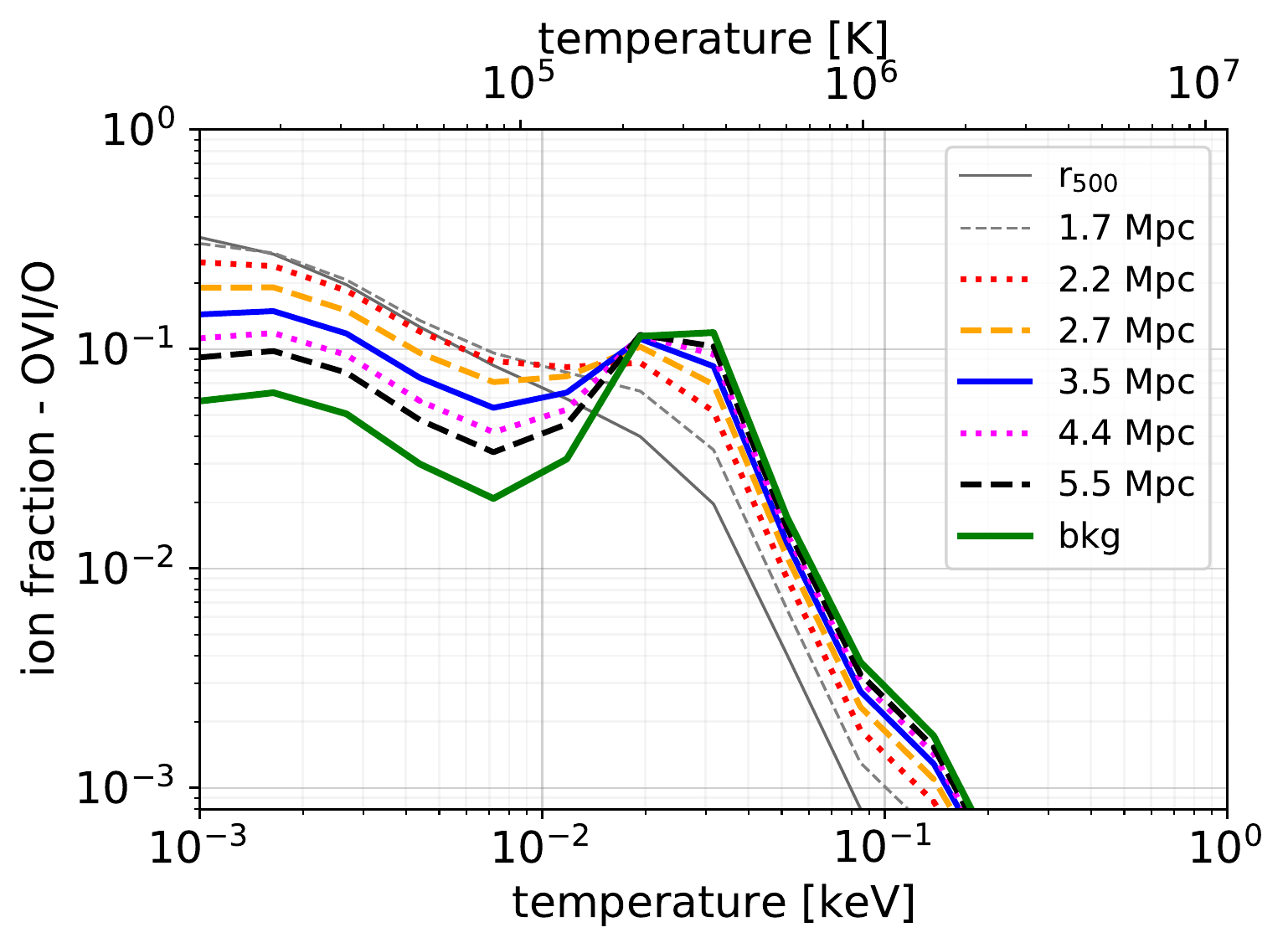}
		\includegraphics{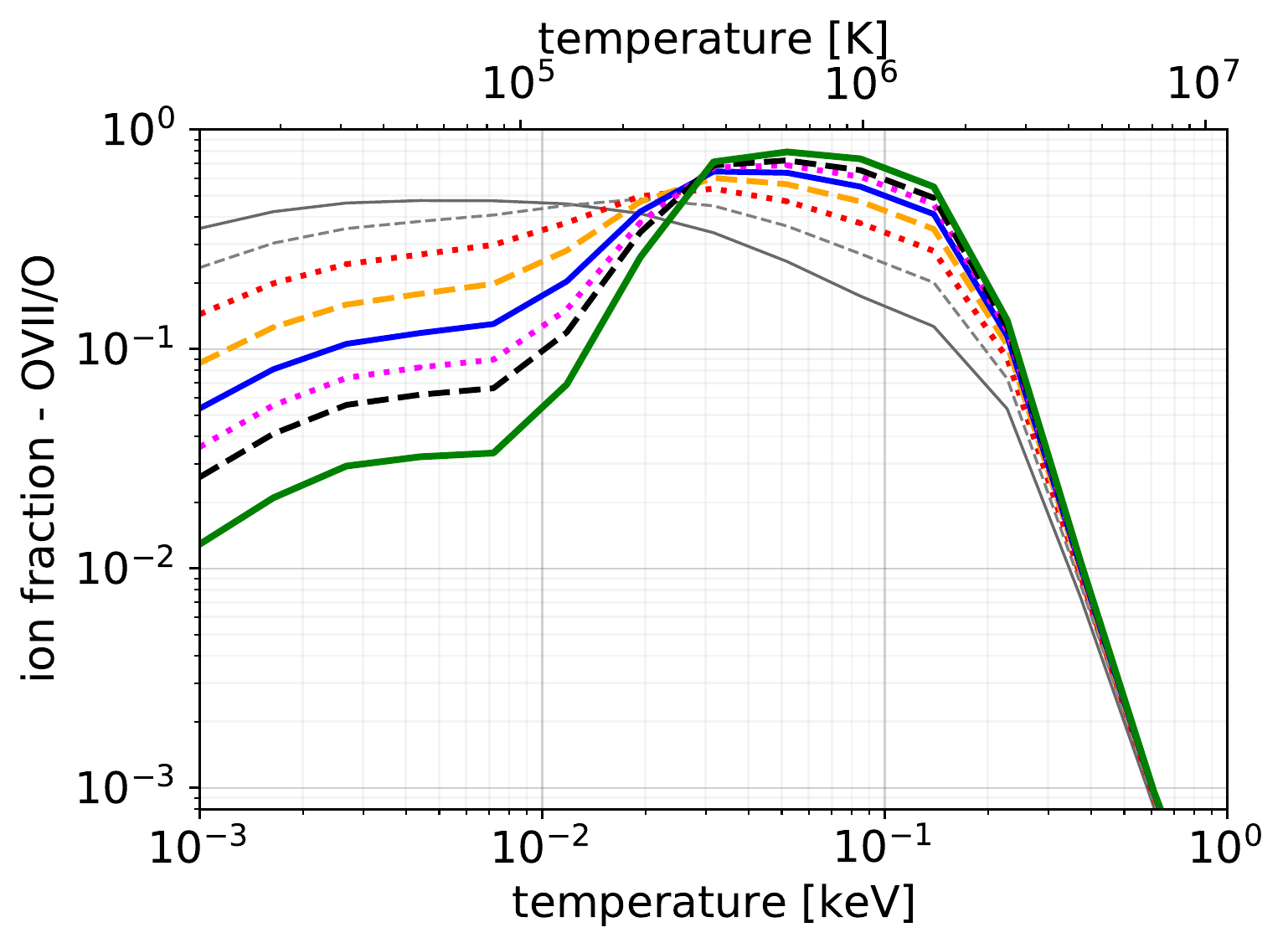} }
	\resizebox{\textwidth}{!}{
		\includegraphics{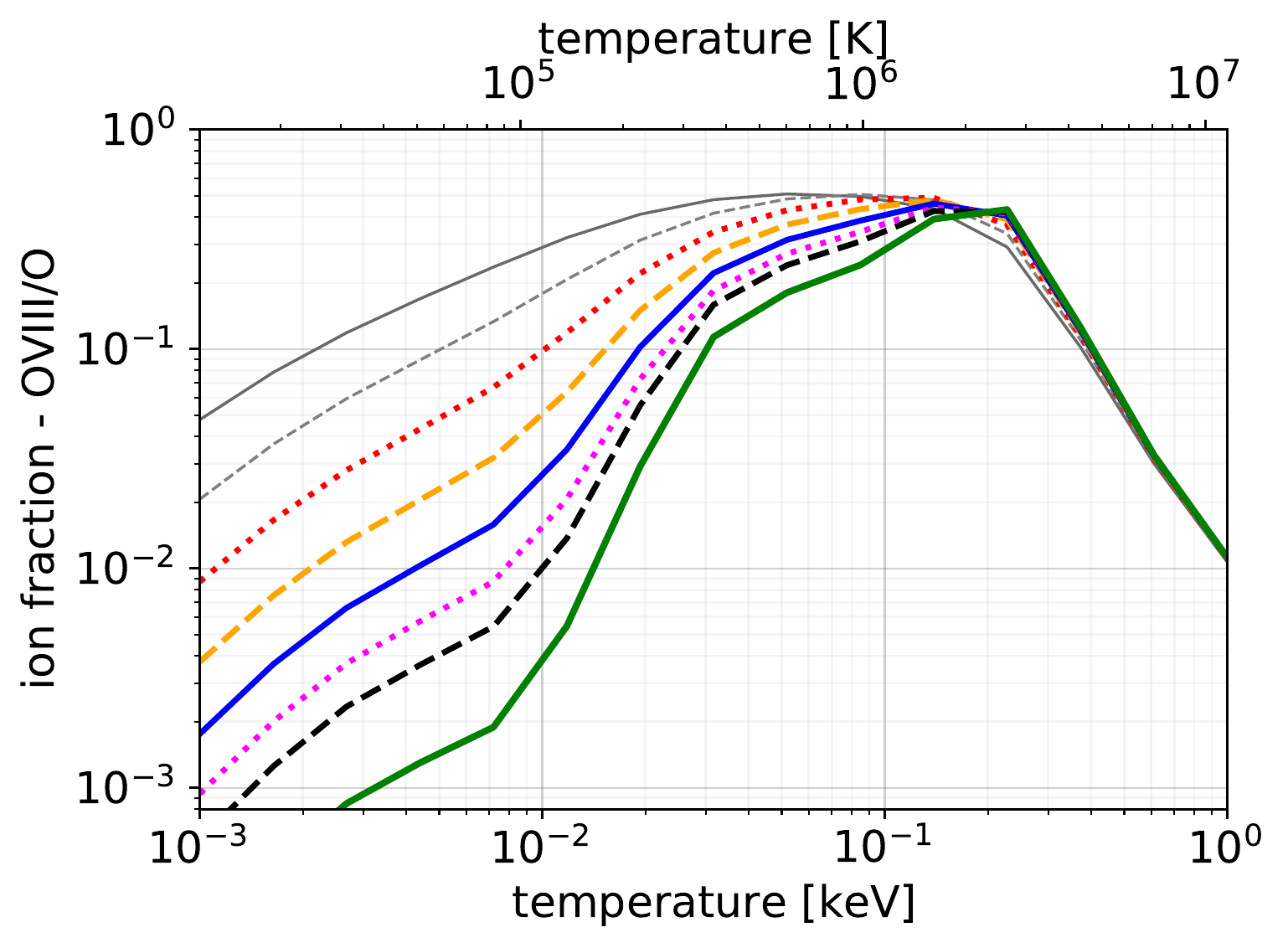}
		\includegraphics{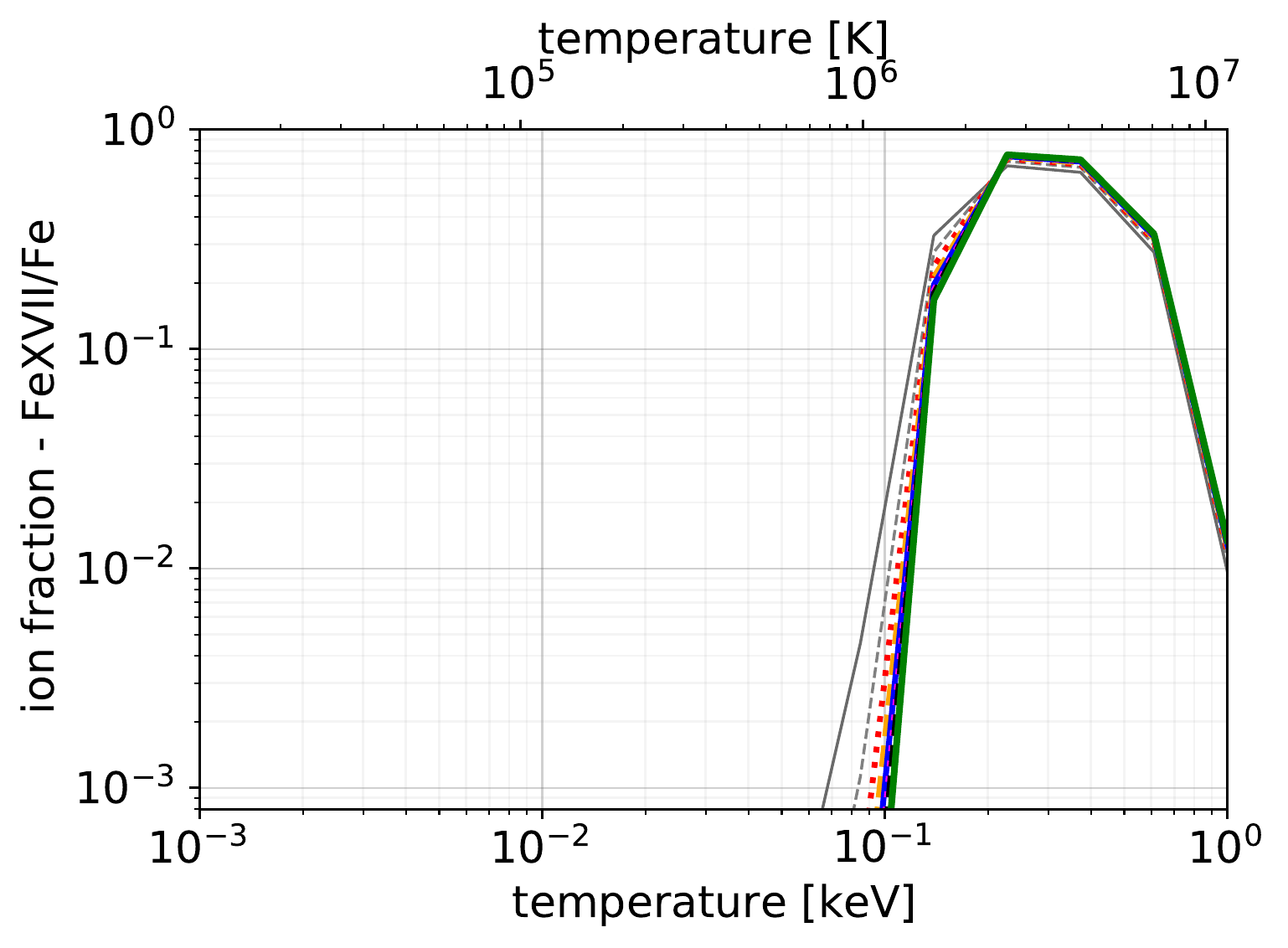} }
	\caption{Same as Fig.\,\ref{Fig:ion_fractions_A2029_update} but for \ion{O}{VI}, \ion{O}{VII}, \ion{O}{VIII} and  \ion{Fe}{XVII}. }
	\label{Fig:ion_fractions_A2029_update_cont}
\end{figure*}

In Figures \ref{Fig:ion_fractions_A2029_update} and \ref{Fig:ion_fractions_A2029_update_cont} we show how ion fractions of different ions (\ion{C}{IV}, \ion{C}{V}, \ion{C}{VI}, \ion{N}{VII}, \ion{Ne}{VIII}, \ion{Ne}{IX}, \ion{O}{VI}, \ion{O}{VII}, \ion{O}{VIII}, \ion{Fe}{XVII}) change as a function of temperature and the distance from the ionisation source (A2029+\emph{bkg}) and compare the results with the ion fractions calculated for the case when the ionisation source is only the UV/X-ray background (\emph{bkg}, green solid line). We plot these ion fractions for density $n_{\rm H} = 2.4 \times 10^{-5}$\,cm$^{-3}$. For each distance in Figures \ref{Fig:ion_fractions_A2029_update} and \ref{Fig:ion_fractions_A2029_update_cont}, Table \ref{Table:U_xil_par} lists a value of the ionisation parameter $\xi$ as well as the dimensionless ionisation parameter for hydrogen $U$, defined as the ratio of the ionising photon flux (in photons per unit area per unit time) to $cn_{\rm H}$, where $c$ is the speed of light and $n_{\rm H}$ is the total hydrogen number density \citep{1979RvMP...51..715D}. In SPEX this parameter is calculated from the SED and the ionisation parameter $\xi$.

\begin{table}
	\centering
	\caption{Distance $R_{\rm abs}$, ionisation parameter $\xi$, which is defined in Eq.\eqref{eq:ionisation_par_xi}, and ionisation parameter for hydrogen $U$ (defined in Sec.\,\ref{sec:ion_balance}).}
	\begin{tabular}{|c|c|c|}
		\hline
		$R_{\rm abs}$ [Mpc] & $\log(\xi)$ [10$^{-9}$ Wm] & U [$\times$ 10$^{-2}$] \\ \hline
		$r_{500}$ & 0.87 & 4.20 \\
		1.7 & 0.70 & 3.59 \\ 
		2.2 & 0.56 & 3.22 \\ 
		2.7 & 0.45 & 3.00 \\ 
		3.5 & 0.36 & 2.86 \\ 
		4.4 & 0.29 & 2.77 \\ 
		5.5 & 0.24 & 2.72 \\ 
		\emph{bkg} & 0.14 & 2.63 \\ \hline
	\end{tabular}
	\label{Table:U_xil_par}
\end{table}

We can see that depending on temperature, the green line representing the results of the background SED (without the contribution of the cluster) forms an upper or lower boundary for the ion fractions. Other lines with different colours represent the ion fractions for the A2029+\emph{bkg} SED for different distances from the cluster centre. In grey we show the distances that are smaller than $r_{200}$ of A2029\footnote{We use the scaling relation $r_{200} \approx \dfrac{3}{2} r_{500}$, which is approximately $2.04$\,Mpc for A2029.}. The temperature at which the background SED fractions transition from a lower to an upper limit is different for different ions. Depending on the temperature we also see that for some ions the differences between the A2029+\emph{bkg} and \emph{bkg} can exceed an order of magnitude (see e.g. \ion{O}{VII}, \ion{O}{VIII}).

\begin{figure*}
	\centering
	\resizebox{\textwidth}{!}{
	\includegraphics{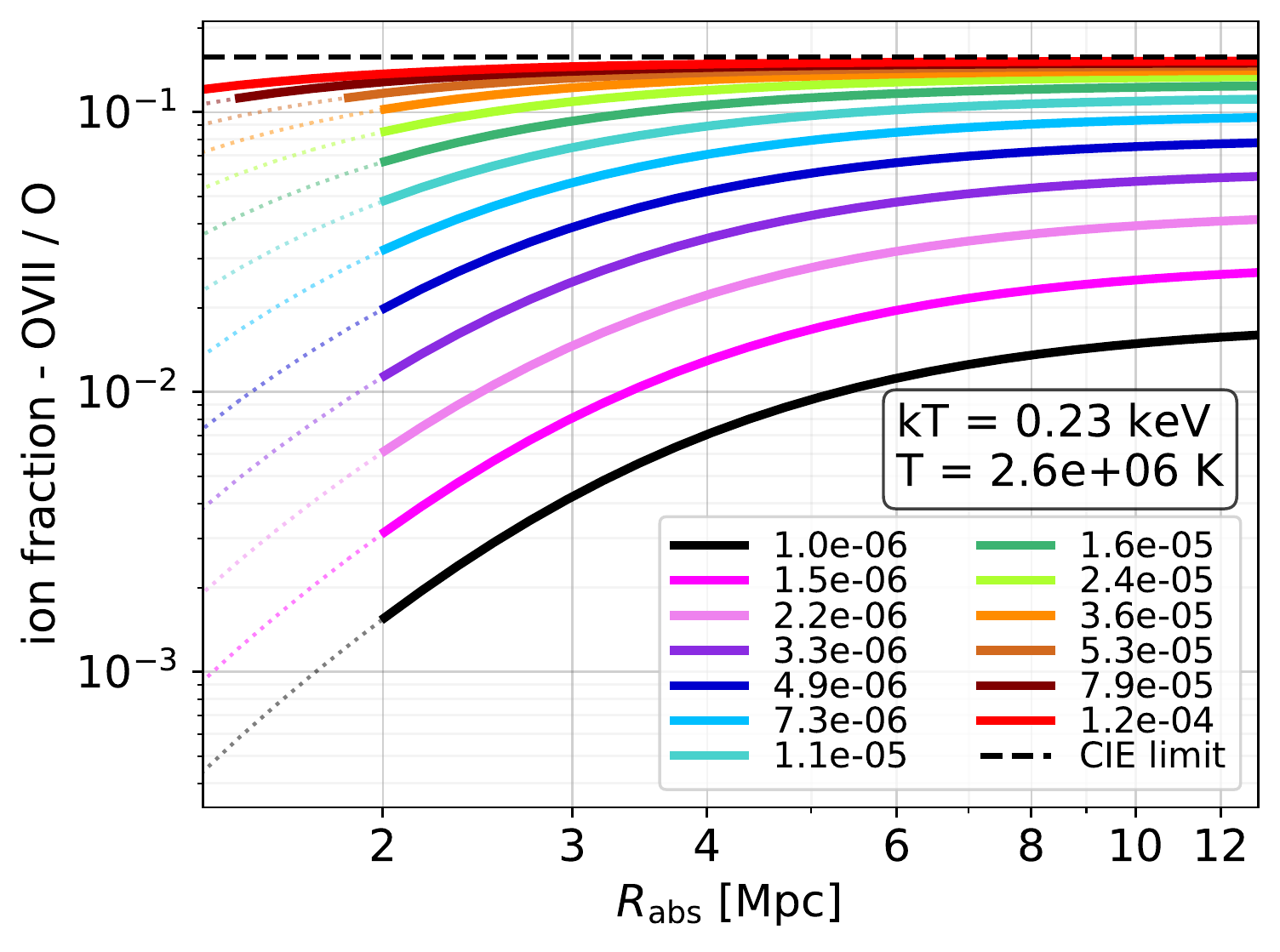}
	\includegraphics{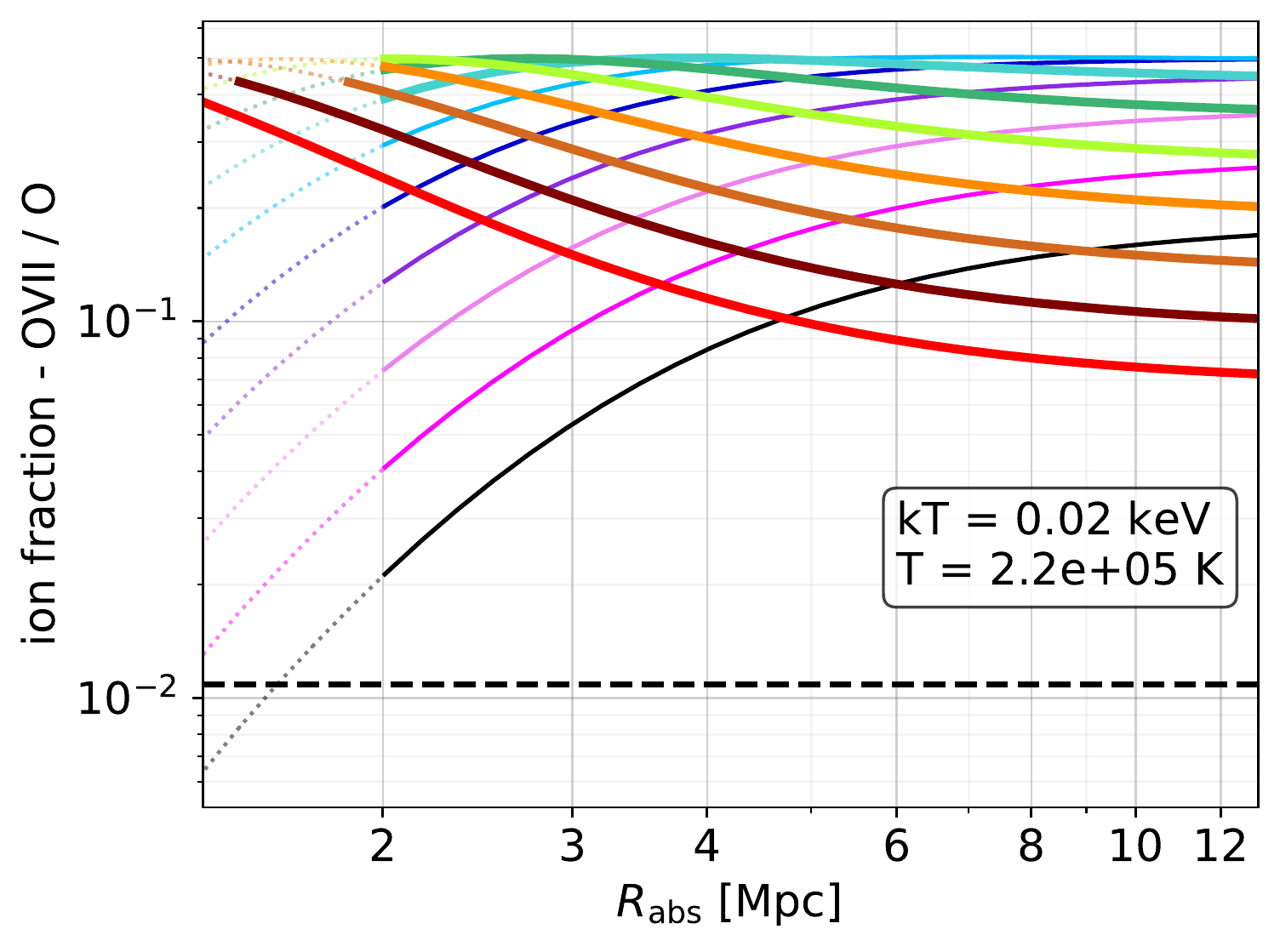}}
	
	\resizebox{\textwidth}{!}{
	\includegraphics{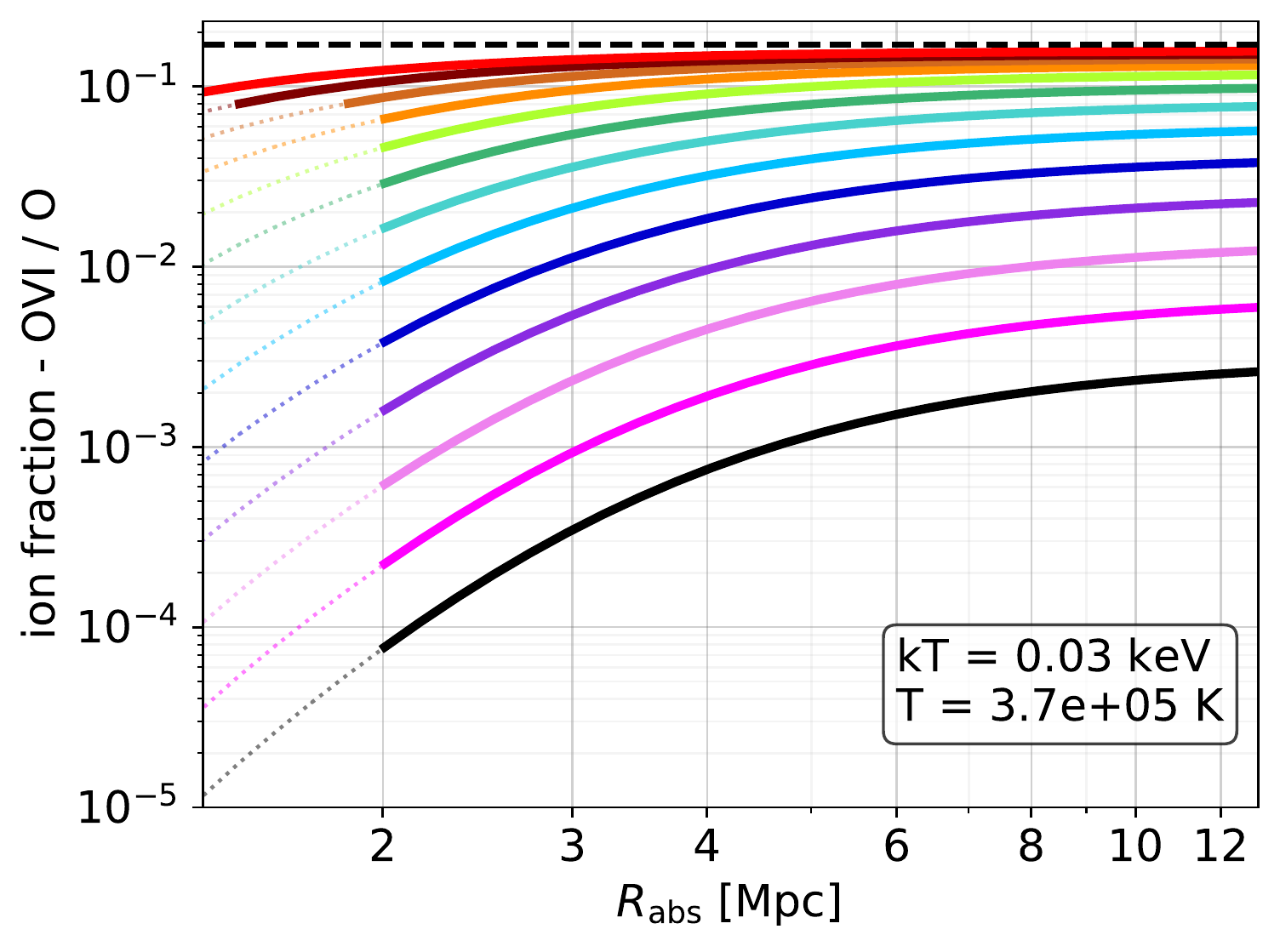}
	\includegraphics{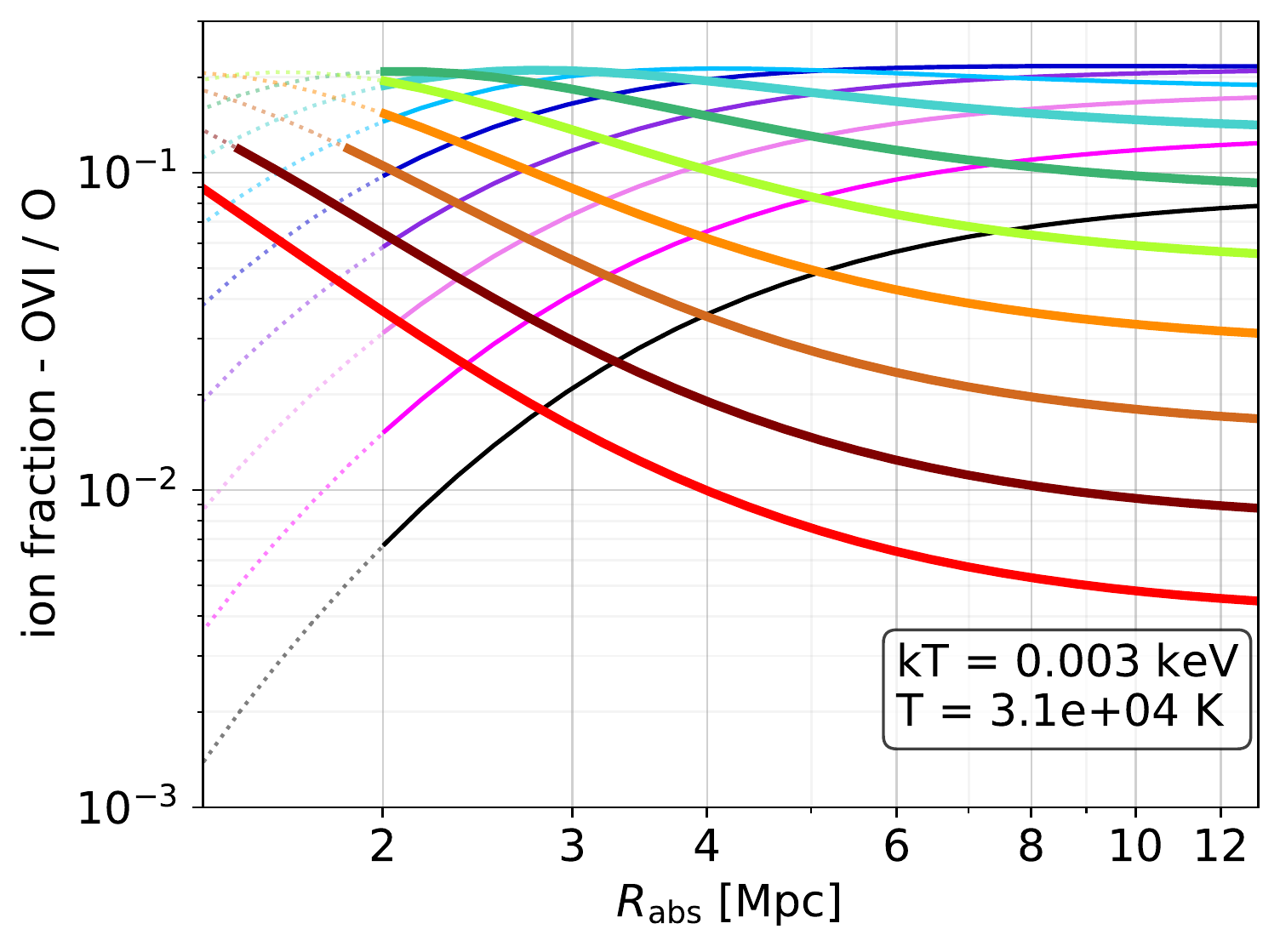}}
	\caption{Ion fractions of \ion{O}{VII} (top row) and \ion{O}{VI} (bottom row) as a function of distance $R_{\rm abs}$ colour coded by the hydrogen number density in units of cm$^{-3}$. Coloured solid lines account for photoionisation by the cluster A$2029$ and the background, the black dashed line shows \ion{O}{VI} and \ion{O}{VII} ion fractions in CIE. Left panels show temperatures, for which the \ion{O}{VI} and \ion{O}{VII} ion fractions peak in CIE, right panels show temperatures 10 times lower. Values that do not satisfy the conditions described in Sec.\,\ref{sec:ion_balance} are shown by dotted lines. }
	\label{Fig:color_bar_plots_main}
\end{figure*}

The ion fractions and the effect of the additional emission from the cluster photons also depend on the density of the plasma as well as the distance from the cluster. Therefore we plot the ion fractions as a function of the distance to the cluster, colour coded by the hydrogen number density in Fig.~\ref{Fig:color_bar_plots_main}, and choose \ion{O}{VI} and \ion{O}{VII} ions as examples. In the collisional ionisation equilibrium (dashed black line), the ion fractions depend only on temperature and do not change with distance, nor density of the gas. 

Fig.~\ref{Fig:color_bar_plots_main} shows the ion fractions of \ion{O}{VI} and \ion{O}{VII} for two different temperatures: the peak temperature in CIE (left panels), and a temperature $10$ times lower (right panels). Coloured solid lines represent the ion fractions of \ion{O}{VI} and \ion{O}{VII} for A2029+\emph{bkg}. All the lines asymptote to the ion fractions for the UV/X-ray background. In this plot, the background would be represented by horizontal lines, where the value would be a constant close to the values where the solid lines flatten towards larger distances. 

One can, however, argue that gas at densities as low as $10^{-6}$\,cm$^{-3}$ is unlikely to exist at a distance of $r_{500}$ from the cluster centre. Therefore, we calculate how the hydrogen number density is expected to change as a function of distance from the cluster using a theoretical curve from \citet{2013MNRAS.432..554W} (their Eq.(20) without scaling factors). This curve can be expressed as 
\begin{equation}
	n_{\rm H}{(r)} = \dfrac{1}{1.2} [P{(r)}]^{3/5} [K{(r)}]^{-3/5}	\,;
	\label{eq:density_distance_eq}
\end{equation}  
where $P{(r)}$ and $K{(r)}$ are the pressure and entropy profiles, respectively. Within $r_{200}$ the cluster is virialized and we can assume that the universal pressure profile $P{(r)}$ follows a generalized Navarro-Frenk-White (GNFW) profile as proposed by \citet{2007ApJ...668....1N} (in this paper we use parameters taken from \citealp{2013A&A...558C...2P}) and that the entropy profile $K{(r)}$ follows a power-law as described in \citet{2010A&A...511A..85P}. For all distances $r \leq r_{200}$ we indicate the densities which fall below the theoretical prediction with dotted lines in Fig.~\ref{Fig:color_bar_plots_main}.

We see that depending on the temperature, the ion fractions can differ significantly (by more than a factor of 10) if cluster emission is taken into account. Depending on the plasma properties and the species, the ion fractions can be higher or lower in comparison with the ionisation by the background. For the temperatures close to the CIE peak, the \emph{bkg} model behaves as an upper limit on the ion fractions, however, if we consider a lower temperature, this only holds true for lower densities while for higher densities the background model acts as a lower limit on the ion fractions instead. Fig.~\ref{Fig:color_bar_plots_main} also shows how the effect of photoionisation by the galaxy cluster emission is stronger for smaller distances from the galaxy cluster and the difference between A$2029$+\emph{bkg} and only the background decreases towards larger distances. In our calculations, the difference between these two models was negligible for distances larger than $13$\,Mpc.

\subsection{Comparison to A262 \& A1795}
\label{sec:comparison_clusters}

\begin{figure}
	\centering
	\includegraphics[width=\columnwidth]{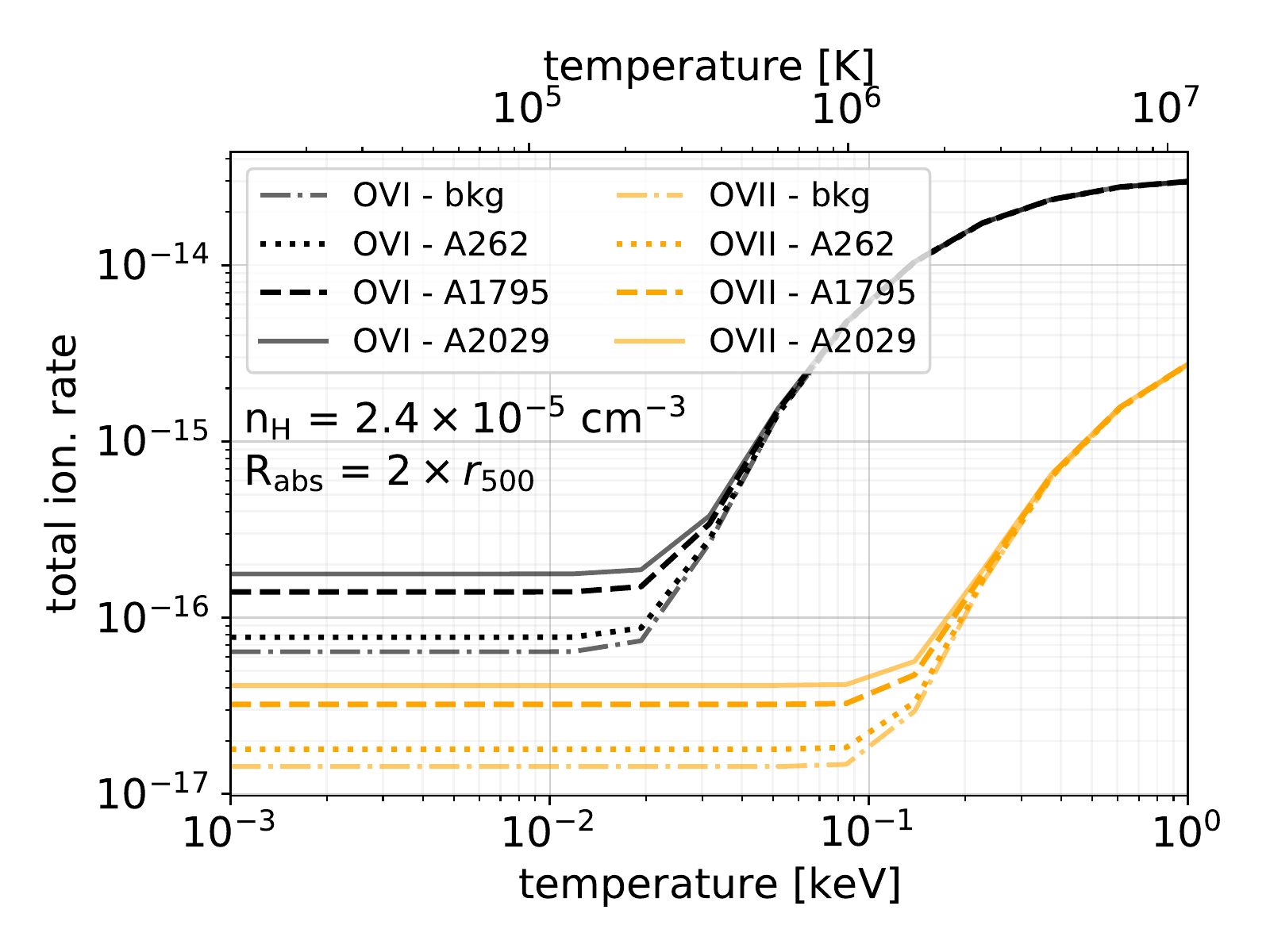}
	\caption{Same as the top panel of Fig.\,\ref{Fig:total_ion_rate_vs_kT} but for ionisation by the background (dash-dotted lines) and by the background plus A2029 (solid lines), A1795+\emph{bkg} (dashed lines) and A262+\emph{bkg} (dotted lines). Black lines show the total ionisation rate for \ion{O}{VI}, orange lines show \ion{O}{VII}. The distance of the photoionised gas from the cluster center is $2\times r_{500}$ and the hydrogen number density is assumed to be $2.4 \times 10^{-5}$\,cm$^{-3}$.}
	\label{Fig:PIE_ion_rate_comparison_all_clusters}
\end{figure}

Fig.\,\ref{Fig:PIE_ion_rate_comparison_all_clusters} shows the same calculation as in the top panel of Fig.\,\ref{Fig:total_ion_rate_vs_kT} for \ion{O}{VI} (black lines) and \ion{O}{VII} (orange lines) with the addition of the less massive clusters A1795 (dashed line) and A262 (dotted line). We see that less massive galaxy clusters have a smaller impact on the photoionization of nearby WHIM compared to more massive clusters.

More quantitatively, let us focus on the  \ion{O}{VII} ion fractions as an example. For A$2029$, the biggest differences between the SED of cluster+\emph{bkg} and \emph{bkg} only are seen for a WHIM temperature $kT = 0.23$\,keV and density $10^{-6}$\,cm$^{-3}$ (black line in Fig.\,\ref{Fig:color_bar_plots_main}, top left panel). In this case, at $r_{200}$ the ion fractions for A2029+\emph{bkg} in comparison with \emph{bkg}-only differ by $162$\%. At the distance of $5$ Mpc, this difference drops to $52$\%, and for distances bigger than $8$\,Mpc the difference is approximately $ 17.4 $\%. If we do the same comparison for A$262$, the differences between ion fractions of \ion{O}{VII} drop to $90$\% (at $r_{200}$), $4.5$\% (at $5$ Mpc) and $3.1$\% (at $8$ Mpc).

In conclusion, the less massive clusters have lower luminosities, and therefore the change in ionisation parameter $\xi$ (between \emph{bkg}-only and cluster+\emph{bkg}) is also smaller. They, however, still alter the ionisation balance and cause a similar effect as we described in Sec.\,\ref{sec:case_study_A2029}. More generalized conclusions are difficult to provide since the ionisation balance is a function of 5 parameters: density and temperature of the photoionised gas, distance of the photoionised gas form the galaxy cluster/group core and the cluster/group mass, and last but not least redshift as well. However, our research suggests that, if WHIM signals are detected in the future in the vicinity of massive clusters of galaxies, the effect of photoioinization from the cluster itself should be modelled carefully.

%%%%%%%%%%%%%%%%%%%%%%%%%%%%%%%%%%%%%%%%%%%%%%%%%%%%%%%%%%%%%%%%%%%%%%%%%%%%%%%%%%%%%%%%%%%%%%%%%%%%
%%%%%%%%%%%%%%%%%%%%%%%%%%%%%%%%%%%%%%%%%%%%%%%%%%%%%%%%%%%%%%%%%%%%%%%%%%%%%%%%%%%%%%%%%%%%%%%%%%%%
\section{Discussion}
\label{sec:discussion}
As we already mentioned in Sec.\,\ref{sec:pion}, due to the addition of the cluster emission, the ionisation balance can no longer be parametrized by the ionisation parameter $\xi$ and the temperature of the photoionised plasma, but it needs to be expressed as a function of the density, temperature, \emph{and} distance of the photoionised plasma to the galaxy cluster. 

In this section we discuss two applications where this behaviour can be demonstrated: firstly, we calculate the column densities of a large scale structure WHIM filament using a toy model in section \ref{sec:column_density_calc} and secondly, we show how the cooling rates are suppressed in section \ref{sec:cooling_rates}.

\subsection{Impact on the predicted column densities}
\label{sec:column_density_calc}

In this section, we consider a filament and calculate the column densities for two different orientations: a) when the line of sight (LoS) is parallel to the spine of the filament, and b) when LoS is perpendicular to the spine of the filament. In both scenarios we use A2029 as a cluster emission source and assume the filament has the shape of a cylinder. We describe the problem in cylindrical coordinates, where $z$ is parallel to the spine of the filament.

In general, we calculate the ionic column density by integrating the ion number density (for a specific orientation) expressed as
\begin{equation}
	n_{i}({r,z}) = \dfrac{\textnormal{Y}_i}{\textnormal{Y}}{(T, n_{\rm H}, R_{\rm abs}(r,z))} \times 0.3 \textnormal{Z}_{\astrosun} \times \rho{(r)} \times \dfrac{\textnormal{X}_{\rm H}}{m_{\rm p}} \;,
	\label{eq:ion_number_ndensity}
\end{equation}
where $\textnormal{Y}_i/\textnormal{Y} = \textnormal{Y}_i/\textnormal{Y}{(T, n_{\rm H}, R_{\rm abs}(r,z))} $ are the ion fractions as calculated in Sec.\,\ref{sec:results} (e.g. \ion{O}{VII}/O), $\rho{(r)}$ is the density of the photoionised plasma, $m_p$ is the mass of a proton and the hydrogen mass fraction X$_{\rm H}$ is set to $0.752$. The metallicity is set to $0.3$ Z$_{\astrosun}$, where Z$_{\astrosun}$ is the solar metallicity taken from \cite{2009LanB...4B..712L} (since we calculate the column density of O, Ne and C, this means that Z$_{\astrosun}$ = [O/H]$_{\astrosun}$ = $10^{8.76}/10^{12}$ for oxygen, Z$_{\astrosun}$ = [Ne/H]$_{\astrosun}$ = $10^{8.05}/10^{12}$ for neon and Z$_{\astrosun}$ = [C/H]$_{\astrosun}$ = $10^{8.39}/10^{12}$ for carbon). For both orientations (parallel and perpendicular), we select two different sets of densities and temperatures according to phase diagrams from the EAGLE cosmological simulations \citep{2015MNRAS.446..521S} shown in \citet{2019MNRAS.488.2947W}: $n_{\rm H}$ = $10^{-6}$\,cm$^{-3}$ and $T = 10^{5}$\,K and $n_{\rm H}$ = $10^{-5}$\,cm$^{-3}$ and $T = 10^{5.5}$\,K. We also consider a density typical for CGM gas: $n_{\rm H}$ = $10^{-4}$\,cm$^{-3}$ and $T = 10^{6.5}$\,K. For simplicity, we set the temperature and density of the filament to constant values in both chosen orientations.

In the first scenario, the LoS is parallel to the spine of the filament and centered on the galaxy cluster core. We calculate the column density (in a pencil beam) of \ion{O}{VI}, \ion{O}{VII}, \ion{O}{VIII}, \ion{C}{V} and \ion{Ne}{VII} by integrating the ion number density $n_i$ as expressed in Eq.\,\eqref{eq:ion_number_ndensity} along the coordinate $z$ from r$_{200}$ to r$_{200}$+$20$\,Mpc, where $20$\,Mpc is our chosen length of the filament. We note that the filament length is still a point of discussion. For example, the maximum filament spine length found in the EAGLE simulation by \citet{2021A&A...646A.156T} is $\approx 35$\,Mpc, while \citet{2020A&A...643L...2T} reports a range of filament lengths between $30$\,Mpc up to $100$\,Mpc, which is not surprising given that the EAGLE simulation box is only $100$\,Mpc on a side. \citet{2020A&A...642A..19M} shows a distribution of filament lengths between $0$--$100$\,Mpc obtained from the galaxy distribution using SDSS measurements. 

In the second scenario, the LoS is perpendicular to the spine of the filament and located at a distance of $z = 2\times r_{500}$ from the center of A2029. In this scenario, we integrate Eq.\,\eqref{eq:ion_number_ndensity} along the $r$ coordinate from zero to the Jeans length $R_{\rm Jeans}$, which is an indicative size for overdense absorbers in any given line of sight \citep{2001ApJ...559..507S}. We set the mean molecular weight to $\mu = 0.625$. Columns $2$ and $3$ in Table\,\ref{tab:column_densities} show the total and ionic column densities for the parallel orientation, columns $4$ and $5$ show the column densities for the perpendicular orientation. Column $6$ shows the column densities for CGM-like properties of the studied gas, equally assuming the depth along the line of sight to be the Jeans length. For each ion we compare the ionic column densities for gas exposed to A2029+\emph{bkg} and to only the background. 

We see that in most of the cases, the addition of the cluster emission to the X-ray/UV background reduces the column densities. The biggest changes are seen for densities $10^{-6}$\,cm$^{-3}$ and $10^{-5}$\,cm$^{-3}$ and can be up to almost factor of $5$ (\ion{O}{VI}). However, in the case of \ion{O}{VIII} and \ion{Ne}{VIII}, we see an enhancement of column densities, which can be $15$--$20$\% for the parallel orientation, and $60$--$70$\% for the perpendicular orientation (for these ions the biggest enhancements are seen for the density $n_{\rm H} = 10^{-5}$\,cm$^{-3}$). 

The results of our toy model show that the changes in column densities are bigger for the perpendicular orientation, and in the case of parallel orientation, these changes do not exceed $50$\%. This is, however, dependent on the length of the filament. If we reduce this length to $10$\,Mpc from $20$\,Mpc, the difference between A2029+\emph{bkg} and \emph{bkg} is almost twice as large. 

The detection of the effect reported in our studies can be challenging with current X-ray missions, however, not impossible. The column densities, currently reported in the literature are approximately of the order of $10^{15}$\,cm$^{-2}$, more specifically, N$_{\ion{O}{VII}}$ = ($1.4 \pm 0.4$)$\times 10^{15}$\,cm$^{-2}$ \citep{2019ApJ...872...83K}; N$_{\ion{O}{VIII}}$ $\sim 9.5 \times 10^{15}$\,cm$^{-2}$ \citep{2002ApJ...572L.127F}; \citet{2007ApJ...665..247W} provides measurements for a set of different ions: N$_{\ion{C}{V}} \sim 10^{15.22}$\,cm$^{-2}$, N$_{\ion{C}{VI}} \sim 10^{15.16}$\,cm$^{-2}$, N$_{\ion{O}{VII}} \sim 10^{16.09}$\,cm$^{-2}$, N$_{\ion{O}{VIII}} \sim 10^{15.80}$\,cm$^{-2}$, N$_{\ion{Ne}{IX}} \sim 10^{15.83}$\,cm$^{-2}$, N$_{\ion{N}{VI}} < 10^{15.39}$\,cm$^{-2}$, N$_{\ion{N}{VII}} < 10^{15.39}$\,cm$^{-2}$, and N$_{\ion{O}{V}} \sim$ ($10^{13.59}$--$10^{14.06}$)\,cm$^{-2}$. \citet{2020A&A...634A.106A} report column densities of \ion{Ne}{IX} and \ion{O}{VIII} to be N$_{\ion{Ne}{IX}} \sim 10^{15.4}$\,cm$^{-2}$ and N$_{\ion{O}{VIII}} \sim 10^{15.5}$\,cm$^{-2}$, respectively. All of these measurements probe WHIM in absorption against bright point-like sources, which would be a suggested method for the detection of the photoionisation of WHIM by galaxy clusters, as reported in this paper. As we can see from Table \ref{tab:column_densities}, our reported column densities that have the biggest differences between cluster+\emph{bkg} and \emph{bkg} are typically of the same order or slightly lower than the column densities currently observed with the X-ray or UV missions. \citet{2022arXiv220315666N} show that the Athena X-IFU will be able to probe absorbing column densities down to N$_{\ion{O}{VII}} \sim 1.6 \times 10^{15}$\,cm$^{-2}$, before problems related to systematic uncertainties on the continuum level become important. A few of the scenarios/geometries listed in our Table \ref{tab:column_densities} can be probed with this limiting sensitivity. However, in order to access the typical column densities where the cluster photoionization has the largest effect (N$_{\ion{O}{VII}} \sim$ a few of $10^{14}$\,cm$^{-2}$, see column $4$ of Table \ref{tab:column_densities}), grating spectrometers with a higher resolving power and which can thus probe lower line equivalent widths robustly, are needed. Missions such as \emph{Arcus} and \emph{Lynx} would therefore be ideal, since their resolving power at $0.5$\,keV is $2500$ and $> 5000$, respectively. Let us take as an example a resonant line of \ion{O}{VII} at $573.95$\,eV. Scaling from Eq.\,(4) and Eq.\,(13) of \citet{2022arXiv220315666N}, for an absorption against a bright Seyfert galaxy with flux of $1$\,mCrab, it would take \emph{Arcus} approximately $500$\,ks and \emph{Lynx/XGS} $50$\,ks to probe column densities of N$_{\ion{O}{VII}}^{\rm A2029 + \emph{bkg}} = 2.75 \times 10^{14}$\,cm$^{-2}$. A detailed feasibility simulation is deferred to future work.

\begin{table*}
	\centering       
	\caption{Total column densities of hydrogen, oxygen, carbon and neon and ionic column densities of \ion{O}{VI}, \ion{O}{VII}, \ion{O}{VIII}, \ion{C}{V}, \ion{Ne}{VIII} for a toy model of a cosmic filament oriented parallel (columns 2 and 3) and perpendicular (columns 4 and 5) at a distance of $2 \times r_{500}$ to the line of sight as described in Sec.\,\ref{sec:column_density_calc} for gas exposed to the SED of the background compared to the SED of A2029+\emph{bkg}. Column $6$ shows the column densities for CGM-like properties of the studied gas, equally assuming the depth along the line of sight to be the Jeans length. For the parallel orientation, the length of the filament is assumed to be $20$\,Mpc, and for the perpendicular orientation its thickness is the Jeans scale. If the column densities for the background and A2029+\emph{bkg} differ by more than a factor of two, we highlight them in the bold face.  }
	\begin{tabular}{|l|c|c|c|c|c}
		\hline
		\textbf{A2029} & \multicolumn{2}{c}{\textbf{parallel orientation}} & \multicolumn{2}{c}{\textbf{perpendicular orientation}} & \textbf{CGM} \\\hline
		$n_{\rm H}$ [cm$^{-3}$] & $10^{-6}$ & $10^{-5}$ &  $10^{-6}$ & $10^{-5}$ & $10^{-4}$  \\
		T [K] &  $10^5$ & $10^{5.5}$ &  $10^5$ & $10^{5.5}$ & $10^{6.5}$   \\ 
		kT [keV] & 0.009 & 0.03 &  0.009 & 0.03 & 0.27  \\
		$R_{\rm Jeans}$ [Mpc] & $3.34$ & $1.88$  &  $3.34$ & $1.88$ & $1.88$ \\
		   \hline \hline
		N$_{\rm H}$ [cm$^{-2}$] & 6.17 $\times 10^{19}$ & 6.17 $\times 10^{20}$ &   1.03 $\times 10^{19}$ & 5.79 $\times 10^{19}$ & 5.79 $\times 10^{20}$   \\ 
		N$_{\rm O}$ [cm$^{-2}$] & 1.07 $\times 10^{16}$ & 1.07 $\times 10^{17}$ &   1.78 $\times 10^{15}$ & 1.00 $\times 10^{16}$ & 1.00 $\times 10^{17}$  \\
		N$_{\rm C}$ [cm$^{-2}$] & 4.54 $\times 10^{15}$ & 4.54 $\times 10^{16}$ &    7.59 $\times 10^{14}$ & 4.27 $\times 10^{15}$ & 4.27 $\times 10^{16}$  \\
		N$_{\rm Ne}$ [cm$^{-2}$] & 2.08 $\times 10^{15}$ & 2.08 $\times 10^{16}$ &   3.47 $\times 10^{14}$ & 1.95 $\times 10^{15}$ & 1.95 $\times 10^{16}$  \\
		\hline \hline
		
		N$_{\ion{O}{VI}}^{\rm A2029 + \emph{bkg}}$ [cm$^{-2}$] & 2.33 $\times 10^{14}$ & 7.64 $\times 10^{15}$ &   \textbf{1.10} $\times 10^{13}$ & \textbf{3.42} $\times 10^{14}$ & 4.11 $\times 10^{13}$  \\  
		N$_{\ion{O}{VI}}^{\emph{bkg}}$ [cm$^{-2}$]  &2.95 $\times 10^{14}$ & 8.68 $\times 10^{15}$ &  \textbf{4.93} $\times 10^{13}$ &\textbf{8.15} $\times 10^{14}$  & 4.24 $\times 10^{13}$  \\  \hline 
		
		N$_{\ion{O}{VII}}^{\rm A2029 + \emph{bkg}}$ [cm$^{-2}$] & 3.19 $\times 10^{15}$  & 6.47 $\times 10^{16}$ &   \textbf{2.75} $\times 10^{14}$ & 4.48 $\times 10^{15}$ & 1.45 $\times 10^{16}$  \\
		N$_{\ion{O}{VII}}^{\rm \emph{bkg}}$ [cm$^{-2}$] & 3.67 $\times 10^{15}$ & 6.85 $\times 10^{16}$ &  \textbf{6.13} $\times 10^{14}$  & 6.43 $\times 10^{15}$ & 1.52 $\times 10^{16}$   \\ \hline
		
		N$_{\ion{O}{VIII}}^{\rm A2029 + \emph{bkg}}$ [cm$^{-2}$] & 5.02 $\times 10^{15}$ & 2.81 $\times 10^{16}$ &    8.44 $\times 10^{14}$ &  4.18 $\times 10^{15}$ & 4.44 $\times 10^{16}$   \\
		N$_{\ion{O}{VIII}}^{\rm \emph{bkg}}$ [cm$^{-2}$] & 4.92 $\times 10^{15}$ & 2.41 $\times 10^{16}$ &   8.22 $\times 10^{14}$ & 2.26 $\times 10^{15}$ & 4.53 $\times 10^{16}$  \\ \hline
		
		N$_{\ion{C}{V}}^{\rm A2029 + \emph{bkg}}$ [cm$^{-2}$]& 2.17 $\times 10^{14}$ & 1.13 $\times 10^{16}$ &    \textbf{1.53} $\times 10^{13}$ & \textbf{0.43} $\times 10^{15}$ & 4.50 $\times 10^{13}$   \\
		N$_{\ion{C}{V}}^{\rm \emph{bkg}}$ [cm$^{-2}$] & 2.72 $\times 10^{14}$ & 1.33 $\times 10^{16}$ &   \textbf{4.54} $\times 10^{13}$ & \textbf{1.25} $\times 10^{15}$ & 4.58 $\times 10^{13}$   \\ \hline
		
		N$_{\ion{Ne}{VIII}}^{\rm A2029 + \emph{bkg}}$ [cm$^{-2}$] & 2.30 $\times 10^{14}$ & 11.35 $\times 10^{14}$ &  2.49 $\times 10^{13}$ & \textbf{1.79} $\times 10^{14}$ & 1.37 $\times 10^{14}$  \\
		N$_{\ion{Ne}{VIII}}^{\rm \emph{bkg}}$ [cm$^{-2}$] & 2.56 $\times 10^{14}$ & 9.33 $\times 10^{14}$ &  4.28 $\times 10^{13}$ & \textbf{0.88} $\times 10^{14}$ & 1.44 $\times 10^{14}$   \\ \hline
	\end{tabular}
	\label{tab:column_densities}
\end{table*}

\subsection{Suppression of cooling rates}
\label{sec:cooling_rates}
In the case of CIE, when no external radiation is present, the cooling rates depend only on the temperature of the CIE plasma, but do not depend on its density after normalisation by $n_e n_{\rm H}$. However, in the presence of the cluster, as we have shown in the previous sections, the normalised cooling rates also depend on a distance $R_{\rm abs}$ and a density $n_{\rm H}$.

\begin{figure}
	\centering
	\includegraphics[width=\columnwidth]{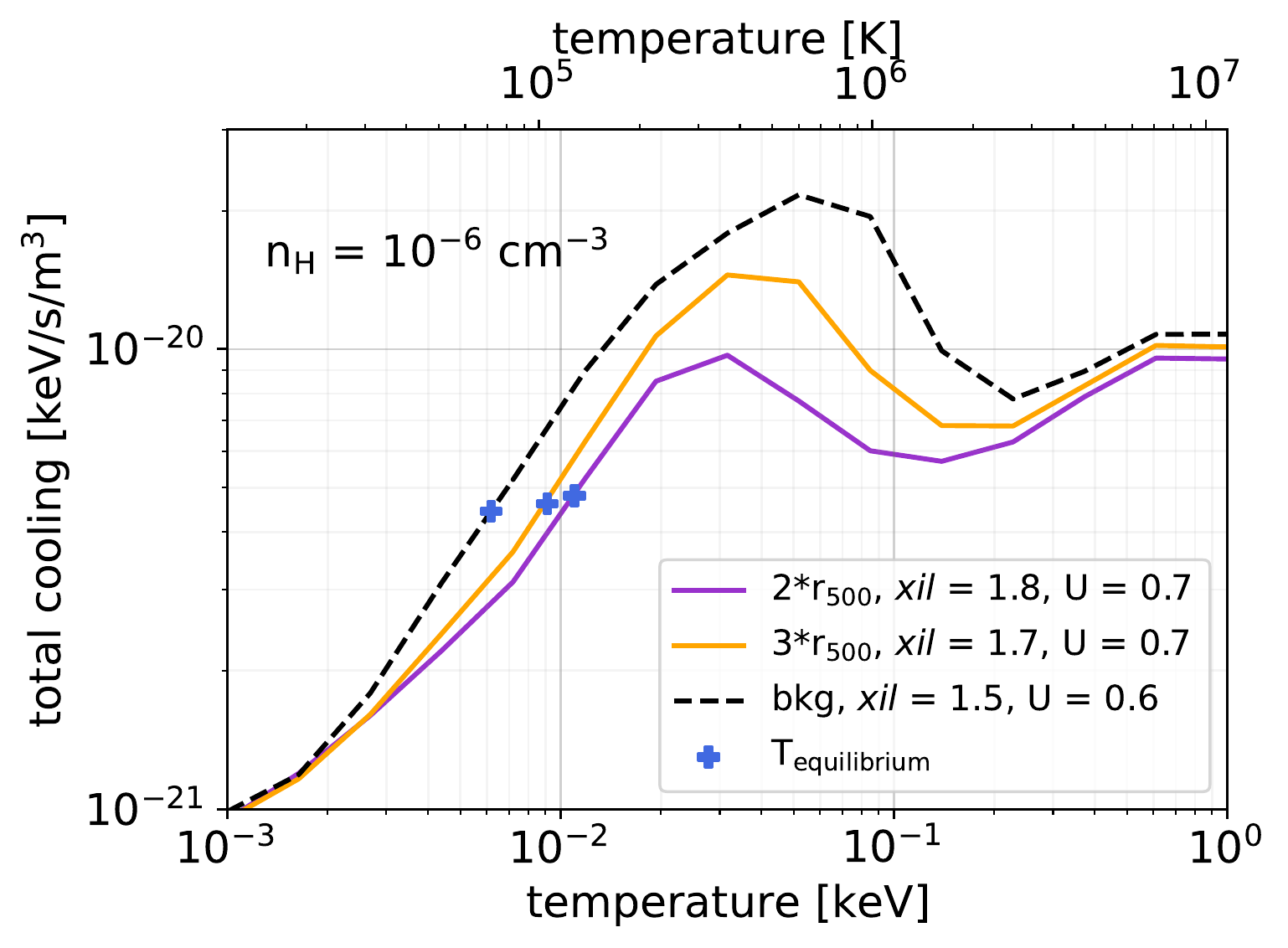}
	\caption{\textit{Left panel:} total cooling rate for the gas with density $n_{\rm H} = 10^{-6}$\,cm$^{-3}$ at distance $R_{\rm abs} = 2 \times r_{500}$ (purple solid line) and $R_{\rm abs} = 3 \times r_{500}$ (orange solid line). The dashed black line represents the total cooling rate for the SED of the background. The equilibrium temperature for each curve is marked by the blue plus sign. Parameter xil is defined as $\emph{xil} = \log{{\xi}}$ in units of $10^{-9}$ Wm. The metallicity is set to $0.3$ Z$_{\astrosun}$ for carbon up to nickel.}
	\label{Fig:cooling_rates_PIE}
\end{figure}

The presence of the photoionising radiation causes a suppression of the cooling rates in comparison with the CIE case (as described in Sec.\,\ref{sec:intro}). In Fig.\,\ref{Fig:cooling_rates_PIE} we show the total cooling in keV/s/m$^3$ for the background SED (black dashed line) and for the SED of A2029+\emph{bkg} for two different distances: $2\times r_{500}$ (purple solid line) and $3 \times r_{500}$ (orange solid line) for density $n_{\rm H}$ = $10^{-6}$\,cm$^{-3}$ and metallicities of $0.3 \times$\,Z$_{\astrosun}$, where similarly to Sec.\,\ref{sec:column_density_calc}, Z$_{\astrosun}$ is taken from \cite{2009LanB...4B..712L}.  As expected, the total cooling rates are suppressed in comparison with only the background, and the closer we get to the cluster, the lower the total cooling rate is. The biggest differences can be seen for temperatures of a few times $10^{-3}$\,keV to $0.2$\,keV. The total cooling rates change at most by a factor of $3$ for temperatures around $0.09$\,keV and densities around $10^{-6}$\,cm$^{-3}$. When we check individual contributions to the total cooling, the biggest change between background and A$2029$+\emph{bkg} is in dielectronic recombination by almost a factor of $7$ (at $kT = 0.05 $\,keV) and for collisional excitation by a factor of $6$ (at $kT = 0.09 $\,keV).

Even though we see that the cooling rates are suppressed once the cluster emission is taken into account, it is also important to calculate on what time scales such cooling happens and whether this change could be relevant. For that we calculate the cooling time $t_{\rm cool}$ for the two most prominent cases from our column density calculations in Sec.\,\ref{sec:column_density_calc}: a) $n_{\rm H}$ = $10^{-6}$\,cm$^{-3}$, T = $10^{5}$\,K, and b) $n_{\rm H}$ = $10^{-5}$\,cm$^{-3}$, T = $10^{5.5}$\,K. The cooling time $t_{\rm cool}$ is expressed as
\begin{equation}
 	t_{\rm cool} =  \dfrac{3 n_{\rm tot} k_{\rm B} T}{2n_{\rm tot}^2 \Lambda}  \;,
 	\label{eq:cooling_time}
\end{equation}
where $n_{\rm tot} = n_i + n_e$ is the total particle number density of gas consisting of the sum of the ion number density $n_i$ and the electron number density $n_e$, and $\Lambda$ is the normalized cooling rate in keV\,m$^3$\,s$^{-1}$. The denominator in Eq.\,\eqref{eq:cooling_time} $n_{\rm tot}^2 \Lambda$ then represents the total cooling rate of the gas in keV\,m$^{-3}$\,s$^{-1}$ which we plot along the $y$-axis of Fig.\,\ref{Fig:cooling_rates_PIE}.

In the case a) we obtain $t_{\rm cool} = 255.5$\,Gyr, and in the case b) $t_{\rm cool}=10.1$\,Gyr. By comparing $t_{\rm cool}$ to the Hubble time ($\approx 14.4$\,Gyr), we see that the suppression of cooling rates will not be important for the gas with density $10^{-6}$\,cm$^{-3}$, but can potentially affect gas with density of $10^{-5}$\,cm$^{-3}$, when the cooling time is around $10$\,Gyr. 
We note, however, that in our calculations we do not include adiabatic cooling due to the Hubble expansion. The addition of cluster emission causes an increase in the thermal equilibrium temperature of approximately $38$\% ($3 \times r_{500}$) or $56$\% ($2 \times r_{500}$) for $n_{\rm H}$ = $10^{-6}$\,cm$^{-3}$ in comparison with the \emph{bkg} model.

If we increase the metallicity to Z$_{\astrosun}$ instead, which might be relevant for the CGM, the biggest differences can be seen again for temperatures of a few times $10^{-3}$\,keV to $0.2$\,keV. The total cooling rates change by a factor of $5$ for temperatures around $0.09$\,keV and densities around $10^{-6}$\,cm$^{-3}$. When we check individual contributions to the total cooling, the biggest change between cooling for the \emph{bkg} model and A2029+\emph{bkg} is in dielectronic recombination and collisional excitation by almost factor of $7$ for both processes at the same temperature as for the metallicity $0.3 \times$\,Z$_{\astrosun}$. The addition of cluster emission causes an increase in equilibrium temperature of approximately $67$\% ($3 \times r_{500}$) or $80$\% ($2 \times r_{500}$) for $n_{\rm H}$ = $10^{-6}$\,cm$^{-3}$ in comparison with the \emph{bkg} model.

%%%%%%%%%%%%%%%%%%%%%%%%%%%%%%%%%%%%%%%%%%%%%%%%%%%%%%%%%%%%%%%%%%%%%%%%%%%%%%%%%%%%%%%%%%%%%%%%%%%%
\section{Conclusions}
\label{sec:conclusions}
This paper studies the impact of the radiation emitted by the intra-cluster gas in galaxy clusters on their environment. In particular, we show how the addition of photons from galaxy clusters alters the ionisation balance of the surrounding warm-hot intergalactic medium relative to models considering any photoionisation by the cosmic UV and X-ray background. 

We model the spectra of the intra-cluster gas from three different cool-core clusters (A$262$, A$1795$, and A$2029$) using the SPEX software package. These spectra, together with the emission from the cosmic UV/X-ray background, are used as ionising sources for the photoionisation model to realistically describe the WHIM. We examine the effect on the ionisation balance for a set of densities and temperatures of the WHIM gas, as well as distances of the WHIM to the galaxy clusters (Sec.\,\ref{sec:results}). In Sec.\,\ref{sec:discussion} we discuss the effect on the column densities for a toy model WHIM filament and calculate how much the cooling rates are suppressed by adding the galaxy cluster to the ionising source. 

Due to the changing spectrum of the radiation that the absorbing gas receives at different distances from the galaxy cluster, the ionisation balance can no longer be solely described as a function of the ionisation parameter $\xi$ and the temperature of the photoionised gas. Instead, the ionisation balance needs to be parametrised as a function of the temperature and density of the photoionised gas, as well as the distance to the galaxy cluster (Fig.\,\ref{Fig:ion_frac_vs_xi_A2029}). 

We see that more massive clusters alter the ionisation balance of the plasma in their vicinity more than the less massive clusters and cause bigger differences in the total ionisation rate (Fig.\,\ref{Fig:PIE_ion_rate_comparison_all_clusters}).

Our main results can be summarised as follows:
\begin{itemize}
	\item For massive, relaxed clusters such as A2029, the addition of the galaxy cluster emission to the UV and X-ray background emission increases the total ionisation rate, especially in the regime of lower densities and temperatures, where photoionisation dominates over collisional ionisation (Figs.\,\ref{Fig:total_ion_rate_vs_kT} and \ref{Fig:total_ion_rate_vs_nH}). 
	\item The ion fractions obtained from the photoionisation by the cosmic UV and X-ray background represent either an upper or lower limit (depending on the plasma properties) on the ion fractions calculated as a function of distance using the cluster+\emph{bkg} emission as opposed to \emph{bkg}-only emission (Figs.\,\ref{Fig:ion_fractions_A2029_update} and \ref{Fig:ion_fractions_A2029_update_cont}).
	\item The effect of the photoionisation by cluster+\emph{bkg} emission is strongest towards the galaxy cluster outskirts and decreases at larger distances from the cluster (Fig.\,\ref{Fig:color_bar_plots_main}). The differences between \emph{bkg} and cluster+\emph{bkg} are negligible for distances larger than $13$\,Mpc. 
	\item The addition of the galaxy cluster emission affects the column densities of our toy filament. For lines of sight passing close to the cluster outskirts, \ion{O}{VI} can be suppressed by a factor of up to $4.5$ ($n_{\rm H}$ = $10^{-6}$\,cm$^{-3}$) or a factor of $2.4$ ($n_{\rm H}$ = $10^{-5}$\,cm$^{-3}$), \ion{O}{VII} by a factor of $2.2$ ($n_{\rm H}$ = $10^{-6}$\,cm$^{-3}$), \ion{C}{V} by a factor of $3$ (for both densities $10^{-6}$ and $10^{-5}$\,cm$^{-3}$), and \ion{Ne}{VIII} can be boosted by a factor of $2$ ($n_{\rm H}$ = $10^{-5}$\,cm$^{-3}$) (Sec.\,\ref{sec:column_density_calc} and Table \ref{tab:column_densities}).
	\item The addition of the cluster emission to the model suppresses the total cooling rates at maximum by a factor of $3$ for the metallicities $0.3 \times $Z$_{\astrosun}$, and by a factor of $5$ for Z$_{\astrosun}$. In both cases, this change is most significant for temperatures ranging from a few times $10^{-3}$\,keV to $0.2$\,keV (for a gas with $n_{\rm H}$ = $10^{-6}$\,cm$^{-3}$).
\end{itemize}

In conclusion, our work emphasises that the impact of the cluster photoionization on the column densities should be considered when interpreting future detections of the IGM near galaxy clusters in absorption.

\vspace{-1.5\baselineskip} 
\section*{Acknowledgements}
The authors acknowledge the financial support from NOVA, the Netherlands Research School for Astronomy. A.S. is supported by the Women In Science Excel (WISE) programme of the Netherlands Organisation for Scientific Research (NWO), and acknowledges the Kavli IPMU for the continued hospitality. SRON Netherlands Institute for Space Research is supported financially by NWO.

%%%%%%%%%%%%%%%%%%%%%%%%%%%%%%%%%%%%%%%%%%%%%%%%%%
\vspace{-1.5\baselineskip} 
\section*{Data Availability}
The dataset generated and analysed during this study is available in the ZENODO repository \citep{lydia_stofanova_2022_6656840}.


\vspace{-1.5\baselineskip} 
\bibliographystyle{mnras}
\bibliography{bibliography} % if your bibtex file is called example.bib

%\begin{thebibliography}{99}
%\end{thebibliography}

%%%%%%%%%%%%%%%%%%%%%%%%%%%%%%%%%%%%%%%%%%%%%%%%%%

%%%%%%%%%%%%%%%%% APPENDICES %%%%%%%%%%%%%%%%%%%%%

\appendix

% Don't change these lines
\bsp	% typesetting comment
\label{lastpage}
\end{document}